\documentclass[pra,
twocolumn,
superscriptaddress,papersize=a4paper,floatfix,longbibliography]{revtex4-2}

\usepackage{amsmath}
\usepackage{graphicx}
\usepackage{bm}
\usepackage{xcolor}
\usepackage{comment}
\graphicspath{{pictures/}}

\definecolor{darkblue}{rgb}{0.1,0.2,0.6}
\definecolor{darkred}{rgb}{0.8,0.1,0.2}
\definecolor{darkgreen}{rgb}{0,0.6,0.1}
\usepackage[colorlinks,citecolor=darkblue,linkcolor=darkred,urlcolor=darkblue]
{hyperref}
\usepackage[all]{hypcap}
\usepackage[normalem]{ulem}
\usepackage{etoolbox}
\usepackage{mathtools}
\usepackage{amsmath}%
\usepackage{amsfonts}%
\usepackage{amssymb}
\newcommand{\bg}{ \begin{gather} }
\newcommand{\eg}{\end{gather}}
\newcommand{\be}{ \begin{equation} }
\newcommand{\ee}{\end{equation}}
\newcommand{\bea}{ \begin{eqnarray} }
\newcommand{\eea}{\end{eqnarray}}

\def\erf{\mathop{\rm erf}}

\newcommand{\p}{\partial}

\begin{document}

\title{The effect of elastic disorder on single electron transport through a buckled nanotube}

\author{S.S. Evseev}

\affiliation{Moscow Institute for Physics and Technology, 141700 Moscow, Russia}

\affiliation{\hbox{L.~D.~Landau Institute for Theoretical Physics, acad. Semenova av. 1-a, 142432 Chernogolovka, Russia}}

\author{I.S. Burmistrov}

\affiliation{\hbox{L.~D.~Landau Institute for Theoretical Physics, acad. Semenova av. 1-a, 142432 Chernogolovka, Russia}}

\author{K.S. Tikhonov}

\affiliation{Skolkovo Institute of Science and Technology, 143026 Moscow, Russia}

\affiliation{\hbox{L.~D.~Landau Institute for Theoretical Physics, acad. Semenova av. 1-a, 142432 Chernogolovka, Russia}}

\author{V. Yu. Kachorovskii}

\affiliation{Ioffe  Institute, Polytechnicheskaya 26, 194021, St.Petersburg, Russia
St.~Petersburg, Russia}
\affiliation{CENTERA   Laboratories,  Institute  of  High  Pressure  Physics, Polish  Academy  of  Sciences,  01-142,  Warsaw,  Poland}

\begin{abstract}
   We study transport properties of a single electron  transistor based on elastic  nanotube.  
         Assuming  that an  external compressive force  is applied to the nanotube,  
     we  focus on  the vicinity of the Euler buckling instability. 
    We demonstrate that in this regime  
     the transport through  the transistor
    is extremely sensitive to  
    elastic disorder. In particular,  
    built-in curvature (random or regular) leads to 
    the ``elastic curvature blockade'':
       appearance of 
    threshold bias voltage in the $I$-$V$ curve  which
    can be larger  
    than the Coulomb-blockade-induced one.
        In the case of a random curvature,  an additional plateau in dependence of the average current on a bias voltage appears.
\end{abstract}

\maketitle

\section{Introduction}

A global trend of modern electronics is the design of nanodevices with ultra-low power consumption  and   a high level of integration. One of the most attractive candidates for this 
purpose  is a single-electron transistor (SET)   which is a sensitive electronic device based on the Coulomb blockade effect. Key  points  
of fundamental  physics of  SET and its operation 
as  a tunable   nanodevice  have been formulated about 20 years ago   (for review see \cite{Zphys1991,Kastner1992,Grabert1992,vanderWiel2002,Hanson2007}). 
There has recently been an increased interest to this  topic  partially   associated with the discovery of carbon systems with a transverse degree of freedom, such  as suspended nanotubes and graphene. 
The interest is motivated by creation  of   nanoelectromechanical systems (NEMS) which are a class of devices integrating electrical and mechanical functionality on the nanoscale  \cite{knobe02_APL,knobe03_Nature,Nishiguchi2003,
armou02_PRL,gorel98_PRL,pisto05_PRL,usman07_PRB,Pogosov2008,Steele2009,Lassagne2009,Weick2010,Weick2011,Micchi2015}.  
The simplest model of SET-based NEMS is provided by a  harmonic mechanical oscillator coupled to an excess  particle number on the SET island \cite{Armour2004,Koch2005,Koch2006,Mozyrsky2006,Pistolesi2007,Pistolesi2008,Leturcq2009,Weick2010,Weick2011,Micchi2015}.

When suspended elastic nanotube is used as a SET island, coupling between mechanical and charge degrees of freedom provides an additional control via mechanical forces. This coupling can be strongly increased by applying a compressive force driving the nanotube towards the  Euler buckling instability \cite{Euler1744,Landau1970} (see more recent experimental \cite{
Falvo1997,Blanter2003,Carr2003,Carr2005,Roodenburg2009,Bonis2018,Barnard2019,Erbil2020,Wang2021} and theoretical  \cite{Carr2001,Werner2004,Peano2006,Saveliev2006,Weick2010,Benetatos2010,Benetatos2010a,Weick2011,Benetatos2011,Micchi2015,Lee2018} studies of this instability).  Physics of such systems is captured by the model developed in  Refs.~\cite{Weick2010,Weick2011,Micchi2015}. This model  was focused on the study of a SET made of a clean nanotube
in which fluctuations 
arise only due to 
temperature $T$. Such fluctuations lead to  
``elastic blockade'': appearance  of a small 
threshold bias voltage
in the SET $I$-$V$ curve 
at the 
center of Coulomb blockade peak \cite{Weick2010,Weick2011,Micchi2015} (the term elastic blockade was introduced earlier for other materials \cite{Nishiguchi2003,Pogosov2008}).  
 However,  the   extreme sensitivity of the nanomechanical SET to an external  environment  implies that presence of  
 disorder or some built-up deflection from ideal symmetry    can strongly modify the 
elastic blockade. 

Typically, nanotubes are not ideal and, therefore,  
are curved (in a random or regular way)  
in their equilibrium state. 
The effect of  
 such built-in curvature  is  
 manifold.  First of all,  
it   can change the effective  potential well confining electrons in the SET island  and redistribute the electron density.  This effect is weak
 provided that the curvature-induced  electronic potential is small as compared to the Fermi energy. 
Secondly, built-in curvature can lead to an electron scattering. This effect is also weak,
since a typical spacial scale of curvature is much larger than the Fermi wavelength. 
The 
 most significant  effect is  
 related to the presence of elastic degree of freedom. In particular,  as was shown in Refs.~\cite{Benetatos2010,Benetatos2010a,Benetatos2011,Lee2018}, 
 built-in curvature  can strongly affect  buckling transition.

\begin{figure*}[t]
\includegraphics[width=0.95\textwidth]{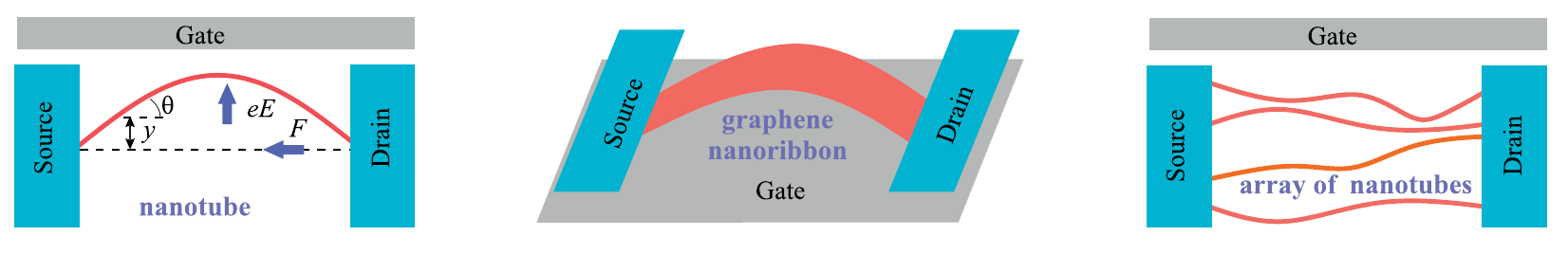}
\caption{\label{fig:2}
 Sketch of a possible realization of SETs with an elastic degree of freedom. Left: A SET based on a suspended  nanotube tunnel-coupled  to a source and a drain and   driven by an external force $F.$ 
A gate electrode creates additional electric 
field $E$ in the $y$ direction. This  transverse field  bends the nanotube provided  it  is occupied by  
an  excessive  electron. Central: A SET with a  graphene nanoribbon buckled by an external force.  Right: A SET made of an array of nanotubes.
}
\end{figure*}

These studies have a great relevance to experiment.     Experimentally, the SETs with elastic degree of freedom  can be realized 
and tested in  various systems, some of which are 
sketched in Fig.~\ref{fig:2}.
The simplest realization is  a  
nanotube located in a plane  and coupled, both electrically and mechanically, with source and drain (see the left panel of Fig.~\ref{fig:2}). Mechanical coupling leads to buckling within the same plane.
Physically,  this system is fully equivalent to a  graphene nanoribbon, which buckles in  the direction perpendicular to plane  (see the central panel in Fig.~\ref{fig:2}). 
Additionally, electro-mechanical coupling  can be   studied   in an array of nanotubes shown in  the right panel of Fig.~\ref{fig:2}.

Recently, a 
possibility of a complete control of   nanotube-based    SETs by means of an external mechanical forces has been clearly demonstrated  experimentally \cite{Bonis2018,Barnard2019,Erbil2020,Wang2021}. 
 In particular, in  
a very recent experimental work \cite{Erbil2020} a double-well elastic potential  describing collective coordinate $Y$   was created  and controlled mechanically by using two mechanical forces:   a compressive force   along the nanotube which leads to buckling bistability and   a force in the perpendicular direction, which makes the  double-well  potential asymmetrical thus ``helping'' the nanotube to choose the proper  potential minimum. This experimental situation  exactly  corresponds to the  theoretical  model we shall focus 
in this paper. In Ref.~\cite{Bonis2018}  the lowest-lying flexural eigenmodes of nanotube  were identified with an unprecedented precision by capacitively driving the  nanotube-based mechanical resonator with external circuit. Excitation of     such eigenmodes by placing the nanotube into the optical cavity  shows  surprisingly high quality factors (up to $10^4!$) \cite{Barnard2019}. The latter   
 proves extreme sensitivity of  suspended nanotubes to external forces.  SET coupled to oscillating field was realized in Ref.~\cite{Wang2021} and  field-driven  Coulomb blockade  peaks were used to  make  a single-electron   
``chopper''.

\begin{figure*}[t]
\includegraphics[width=0.95\textwidth]{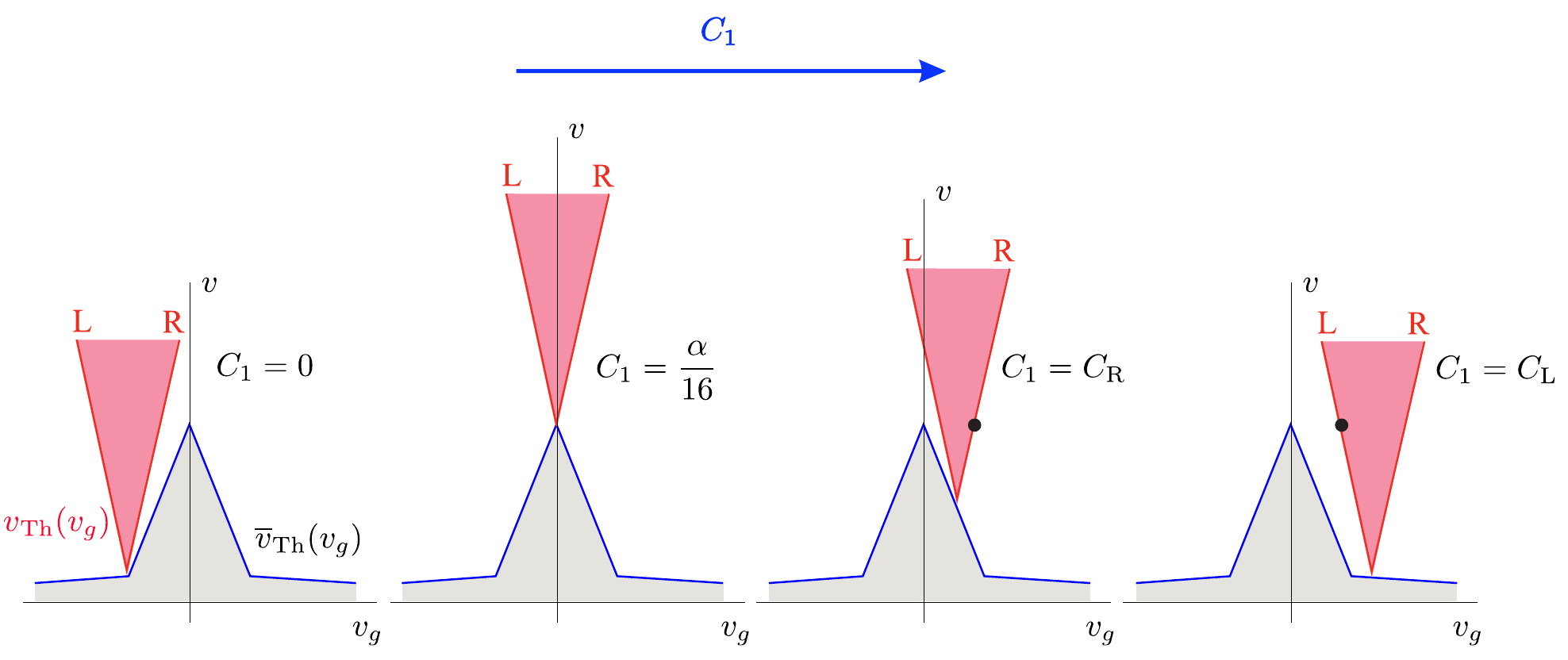}
\caption{\label{fig:1}Schematic plot of  the current density in $(v_g,v)$ plane     
for  different realizations of a  
 built-in  curvature, characterized by disorder strength, $C_1$. Disorder increases from  the left panel  to the  right one. 
 Regions with non-zero current, $v{>}v_{\rm Th}(v_g),$    are marked 
 by pink color. They have a form of 
 triangle  
 with right (R)  and left (L) boundaries described by 
 function $v_{\rm Th}(v_g)$ 
  and shown by red thick lines.     
   Small gap  corresponding to  zero current  in the $I$-$V$ curve in the clean case, $C_1=0,$  is dramatically  increased in the certain interval of disorder strength, and is maximal  for  $C_1{=}\alpha/16$  
 due to the elastic curvature blockade. For any point in  $(v_g,v)$ plane there are two  values of  disorder $C_1{=}C_{\rm R}$   and  $C_1{=}C_{\rm L}$    corresponding to crossing of this point, respectively,  by the right and left boundaries  of the current-carrying 
triangle ($C_{\rm L}{>}C_{\rm R}$).  Both $C_{\rm L} $ and $ C_{\rm R}$ depend on coordinates $v_g,v$, $C_{\rm R,L}{=}C_{\rm R,L}(v_g,v).$ Area where current is zero for any disorder, $v{<}\overline {v}_{\rm Th}(v_g)$  is marked 
by grey color. The function $\overline {v}_{\rm Th}(v_g)$ is shown by thick blue line. 
}
\end{figure*}

 In this paper,  we  study disordered nanotube- or nanoribbon-based   SET under the action of a mechanical force  (see Fig. \ref{fig:2}).  We model disorder by spontaneous local curvature  (regular or random)   \cite{Benetatos2010,Benetatos2010a,Benetatos2011,Lee2018}. 
We  
demonstrate that the results  obtained in Refs.~\cite{Weick2010, Weick2011,Micchi2015} for  
a clean nanomechanical SET
are strongly modified in the presence of 
 such curvature.
This 
built-in curvature   leads --- due to electro-mechanical coupling --- to two 
crucial effects: (i) existence of large 
threshold bias voltage in $I$-$V$ curve at the center of Coulomb blockade peak for a fixed,  
built-in curvature (see Fig.~\ref{fig:1}), 
(ii) appearance of an intermediate 
plateau in bias-voltage dependence of the current 
averaged over realizations of random curvature (see  bottom panel  in Fig.~\ref{fig:6}).

 Outline of the paper is as follows. We start with  formulation of the model in Sec. \ref{Sec:Model}. In Sec. \ref{Sec:FundMode} the approximate treatment of the model is developed. The current for a fixed curvature of nanotube is calculated in Sec. \ref{Sec:fixed}. The case of a random curvature is considered in Sections \ref{Sec:random} and \ref{Sec:theta-PDF}. 
 We end the paper with discussions and conclusions in Sec. \ref{Sec:Disc}. Some technical details are given in Appendices.

\section{Model\label{Sec:Model}}

The sketch of the setup is shown in Fig. \ref{fig:2} (left panel). We consider a nanotube  of length $L$  suspended between left   and   right  
leads, which serve as source and drain, respectively. The tunneling coupling of the  
nanotube to the source (drain) is characterized by the tunneling rates $\Gamma_L$ ($\Gamma_R$). We assume that a finite voltage 
 bias $V$ is applied between source and drain shifting their chemical potentials to ${\pm}eV/2$. The electrical potential $V_{\rm g}$ of the nanotube can be tuned by a capacitively coupled electrode (gate).  
  We consider regime of  
 strong Coulomb blockade, $T{<}E_{\rm c},$  
  such that
 only a  single excessive (in addition to a quasi-neutral Fermi gas) electron  can occupy a nanotube. 
 Here,  $E_c$ is the charging energy ($E_{\rm c} {\sim} 10$  meV   for 
 a typical nanotube  length $L{\sim} 0.1~ \mu$m).
  The  
 Hamiltonian  of a nanotube,  
  including  the elastic degrees of freedom parametrized by the bending angle $\theta$ as a function of the arc-length position $s$
   reads 
\begin{gather}
\hat H = \int_0^L ds \
		\Bigl [ \kappa(\theta^\prime-C^\prime)^2/2 + F \cos \theta\Bigr ]  + eE\hat n_{\rm d} Y .
\label{eq:Ham}		
\end{gather}
Here $\theta^\prime{\equiv} {d\theta}/{ds},$  
$\kappa$ denotes the bending rigidity of the nanotube, $F$ stands for the compressing force applied to the ends of the nanotube,  $E$ is an additional electric field created by the gate electrode, directed along $y$ axis,
and 
\be Y= \int_0^L |\psi_0(s)|^2 y(s) ds 
\label{Y}
\ee 
is the average displacement of the nanotube in $y$ direction. Here, $y(s)$ is transverse coordinate of the nanotube, related to  bending angle as follows,  
\be
\frac{dy}{ds}=\sin \theta .
\label{dyds}
\ee 

We model  disorder by  built-in  curvature  described by  function  $C(s),$ which can be regular or random. The electronic degrees of freedom enters the Hamiltonian \eqref{eq:Ham} via the operator of the excess particle number, $\hat n_{\rm d}$, and  the wave function,  $\psi_0(s)$, of an active 
energy level  for an excessive electron  on the nanotube. Provided the nanotube is occupied by  an
 additional  electron,  the quasi-neutrality is  broken, and  the transverse electric field $E$ bends the nanotube.  Hence, there is the 
term $eE\hat n_{\rm d} Y$  in Eq.~\eqref{eq:Ham}.  The nanotube is assumed to be clamped on 
the left and right leads  
of the SET, such that Hamiltonian \eqref{eq:Ham} has to be supplemented by the following curvature-dependent boundary conditions \cite{Benetatos2010a,Benetatos2010},
\begin{equation}
y(0)=y(L), \quad \theta^\prime(0)=C^\prime(0), \quad 
    \theta^\prime(L)=C^\prime(L) .
     \label{BC:1}
\end{equation}

In the Hamiltonian \eqref{eq:Ham} we neglect the elastic kinetic energy. This is justified under assumption that a typical elastic frequency of transverse oscillations of the nanotube, $\omega {\propto} \sqrt{\kappa/(\rho L^4)}$, where $\rho$ is the linear mass density of the nanotube, is small as compared to temperature,  
$\hbar  \omega {\ll} T$.  As a result, the only dynamical degree of freedom in the Hamiltonian 
\eqref{eq:Ham} is the excess particle number, $\hat n_{\rm d}$. In the strong Coulomb blockade regime, the operator $\hat n_{\rm d}$ takes two values, $n_{\rm d}{=}0$ and $n_{\rm d} {=}1$. Solution of the coupled --- electronic and elastic --- dynamical problem can be essentially simplified in the adiabatic approximation provided the tunneling between the nanotube and the source and the drain is sufficiently fast, 
$\Gamma_{\rm L,R} {\gg} \omega$. In such adiabatic approximation and under assumption,
  $\hbar \Gamma_{\rm R,L}{\ll} \max\{T, |eV|, |eV_g|\}$, 
 the computation of the bending angle becomes fully classical and the bending angle profile is described by the following nonlinear equation (for $C(s){\equiv} 0,$ this equation was derived in Ref.~\cite{Weick2011})
\begin{equation}
\kappa\left[\theta^{\prime\prime}(s)-C^{\prime\prime}(s)\right]\! +\! F \sin \theta(s) = e 
E(s) 
n_{\rm d}(Y) \cos\theta(s) ,
\label{eq-for-theta}
\end{equation}
where $E(s) {=} E \int_s^L ds^\prime |\psi_0(s^\prime)|^2$ and $n_{\rm d}(Y)$
is the average value of the operator $\hat n_{\rm d}$ for a fixed configuration $\theta(s)$,
\begin{equation}
n_{\rm d}(Y) =  
\sum \limits_{\lambda=\pm} \gamma_\lambda f_{\rm F}\bigl(eEY-e V_\lambda 
\bigr) .
\label{eq:ndY}
\end{equation}
Here $f_{\rm F}(\varepsilon){=}1/[\exp(\varepsilon/T){+}1]$ denotes the Fermi function,
$V_\pm=V_{\rm g}\pm V/2$,
$\gamma_-{=}\Gamma_{\rm L}/(\Gamma_{\rm L}{+}\Gamma_{\rm R})$, and  $\gamma_+{=}
\Gamma_{\rm R}/(\Gamma_{\rm L}{+}\Gamma_{\rm R}).$
Under  the same assumptions,  
the SET current 
reads 
\begin{equation}
\label{J}
\mathcal{I} = \frac{2 e \Gamma_R\Gamma_L I(Y)}{\Gamma_L+\Gamma_R}, \quad 
I(Y) {=}\sum \limits_{\lambda{=}\pm} \frac{\lambda}{2} f_{\rm F}\bigl( eEY-eV_{\lambda}\bigr) .
\end{equation}
\color{black}
In order to compute the current by means of  Eq. 
\eqref{J}, one needs to solve 
the nonlinear second order differential Eq. \eqref{eq-for-theta}. Although it can be done numerically, as we shall demonstrate below, one can construct an analytic solution near buckling instability.

\section{Fundamental mode approximation\label{Sec:FundMode}} 

In order to satisfy the boundary conditions \eqref{BC:1}, it is convenient to expand the functions $\theta(s)$ and $C(s)$  in the Fourier series, 
\begin{equation}
\label{expansion}
    \theta(s) = \sum\limits_{n=1}^{\infty}\theta_n \cos q_n s,\quad C(s)= \sum\limits_{n=1}^{\infty} C_n \cos q_n s,
\end{equation}
where 
$q_n {=} \pi n / L$. Euler buckling instability is clearly seen from Eq.~\eqref{eq-for-theta} at $E{=}0$. Indeed, linearization of this equation yields divergence of $\theta_1$, for $F{=}F_{\rm c}$: 
\be 
\theta_1 \approx - \frac{8 C_1}{\epsilon}, \quad \epsilon= \frac{8(F -F_{\rm c})}{F_{\rm c}} , 
\label{theta-lin-C1}
\quad F_{\rm c}{=}\kappa\left(\frac{\pi}{L}\right)^2 ,
\ee
for the critical force of the instability (see Refs. \cite{Euler1744,Landau1970}). 
 
Hereafter we assume that $|\epsilon|{\ll}1$. In fact, the growth of $\theta_1$ for $|\epsilon| {\to} 0$ is limited by nonlinear terms in the Eq. \eqref{eq-for-theta}. The modes $\theta_n$ with $n{\geqslant} 2$ are finite at $F{=}F_{\rm c}$ and can be found within the linear approximation:  $\theta_n {\sim} C_n$.  Such crucial difference between the behavior of the mode with $n{=}1$ and modes with $n{\geqslant}2$ allows us to project Eq. \eqref{eq-for-theta} onto the fundamental mode $\cos(q_1s)$ as it was done in the clean case \cite{Weick2010,Weick2011}.  In the most of our paper, we  throw out all but the first mode (weak effects coming from modes with  $n{\geqslant} 2$ are briefly discussed at the Appendix~\ref{high-harm}). The shape of nanotube, within  fundamental mode approximation can be found from Eq.~\eqref{dyds} and is given by
\be
y_1(s)=\frac{L \theta_1 }{\pi} \sin\left ( \frac{\pi s}{ L}\right). 
\label{y1-s}
\ee

We shall also make several simplifying assumptions throughout the paper.  We 
assume 
the symmetric SET,
$\Gamma_R{=}\Gamma_L{=}\Gamma$ 
(i.e., $\gamma_+{=}\gamma_-{=}1/2$). Also we 
set $T{=}0$. We shall estimate  the effect of thermal fluctuations  at the end of the paper and  demonstrate  that for realistic values of parameters this effect is small. The latter assumption allows one to replace Fermi functions entering Eqs.~\eqref{eq:ndY} and \eqref{J}  with step-functions.
We shall also assume that  disorder is weak  so that    $|\theta_1|{\ll} 1.$  Then, expanding sinus  in  Eq. \eqref{eq-for-theta} as  $\sin \theta {\approx} \theta{-}\theta^3/6,$  keeping  only the first harmonic,  $\theta{\approx} \theta_1 \cos(q_1 s)  ,$ and projecting thus obtained equation  onto the first mode,    we obtain the closed nonlinear  balance equation for the amplitude of the first harmonic  $\theta_1$: 
\begin{equation}
f(\theta_1)=0, 
\label{eq:BA:1}
\end{equation}
where  
\begin{align} 
f(\theta_1) &=\theta_1^3-\epsilon \theta_1+\alpha n_{\rm d}(\theta_1) - 8 C_1,
\label{f-theta}
\\
n_{\rm d}(\theta_1) &=
\frac{1}{2} \left[ \Theta(v_+-\theta_1)+  \Theta(v_--\theta_1)\right]
.
\label{nd-theta}
\end{align}
\color{black} 
   Here  $\Theta(x)$ denotes the Heaviside step function. 
   Within the same approximation the current is given by
  \begin{equation}
\label{J-theta}
\mathcal{I} =  e \Gamma I(\theta_1), \quad
I(\theta_1) {=}\frac{1}{2} \left[ \Theta(v_+{-}\theta_1){-}  \Theta(v_-{-}\theta_1)\right] .
\end{equation}
     In Eqs.~ \eqref{f-theta}, \eqref{nd-theta}, and \eqref{J-theta} we introduced  dimensionless strength of the electron-phonon interaction 
     and the
dimensionless voltages
\cite{footnote}
\begin{equation} 
\begin{aligned}
&\alpha=\frac{16 a_1 eE}{\pi F_{\rm c}}, \quad v_\pm {=} v_{\rm g}{\pm} \frac{v}{2} ,
\\
&
 v_{\rm g}=\frac{\pi V_{\rm g}}{a_1 E L}, \quad v = \frac{\pi V}{a_1 EL}, 
 \end{aligned}
\label{vvg-dimen}
\end{equation}
where  $ a_1 {=} \int_0^L ds\, |\psi_0(s)|^2 \sin(q_1 s) $  is the numerical coefficient  which fully encodes  the dependence  on the wave function  of the active electron quantum level: it changes 
from  $a_1{=}8/3\pi$  for $\psi_0(s){=}\sqrt{2/L} \sin(q_1 s)$    to  $a_1{\approx}2/\pi$  for $\psi_0(s){=}\sqrt{2/L} \sin(q_N s)$ with $N {\approx} k_{\rm F} L/\pi {\gg} 1 $. Here $k_{\rm F}$ stands for the Fermi momentum.  

  Importantly,  the typical  magnitude of $\alpha$ is  very small, $\alpha {\sim} 10^{-4} \div 10^{-2}$ \cite{Weick2011}, that will allow us to study the effect of the electro-mechanical coupling perturbatively.

   We are interested in the solution $\theta_1(C_1)$ of Eq. \eqref{eq:BA:1} that provides the minimum of the  dimensionless  energy 
  \begin{equation}
W(\theta_1)= \frac{(\theta_1^2-\epsilon)^2}{4}  +\alpha w_{\rm d}(\theta_1) - 8 C_1 \theta_1 ,
\label{W-theta1}
\end{equation}
related to $f(\theta_1)$ as follows: $f(\theta_1){=}\p W(\theta_1)/\p \theta_1.$ Here we introduce
\begin{equation}
    w_{\rm d}(\theta_1) = \int_0^{\theta_1}n_{\rm d} (\theta) d\theta =\frac{1}{2}
    \sum_{\lambda=\pm} (\theta_1-v_\lambda)\Theta(v_\lambda-\theta_1) . 
    \label{wd}
\end{equation}
We subtracted  from $w_{\rm d}$  the 
$\theta_1-$independent term 
$[v_+\Theta(v_+){+}v_-\Theta(v_-)]/2$ and  assumed that $v_+{>}v_-$ ($v{>}0$).
We note that in the single mode approximation the  energy, $\mathcal E(\theta_1),$  corresponding to Hamiltonian \eqref{eq:Ham}  
is connected with  the dimensional energy $W(\theta_1)$ as follows 
\be
\mathcal E(\theta_1)= \frac{\pi^2 \kappa W(\theta_1)}{16 L}.
\label{E-W}
\ee

  For $\alpha{=}0$ the energy $W(\theta_1)$ has the form of the Landau expansion of the free energy in series of the order parameter $\theta_1$ near the second order phase transition at $\epsilon{=}0$. The  curvature $C_1$ plays the role of the  
  symmetry-breaking  field that breaks the symmetry between two states of minimal energy at $\epsilon{>}0.$ Hence,   the system can  show the  hysteretic  behavior. However, since we focus here in dc current, we assume that the system always has time to reach the absolute energy minimum.   

 Most importantly, the single-mode approximation allows 
 us to find the $I$-$V$ curve for a 
fixed 
curvature  (i.e. for given $C_1$). For random curvature which is characterized by a certain distribution function $\mathcal P(C_1)$, this approximation allows us to calculate analytically the distribution function for the current and analyze the competition of  the random curvature with  the term $\alpha w_{\rm d}(\theta_1),$  which suppresses the  bistablity. 
Below, we discuss the cases of fixed and random curvature in Sections \ref{Sec:fixed} and \ref{Sec:random}, respectively.

\begin{figure*}[t]
\centerline{\includegraphics[width=0.95\textwidth]{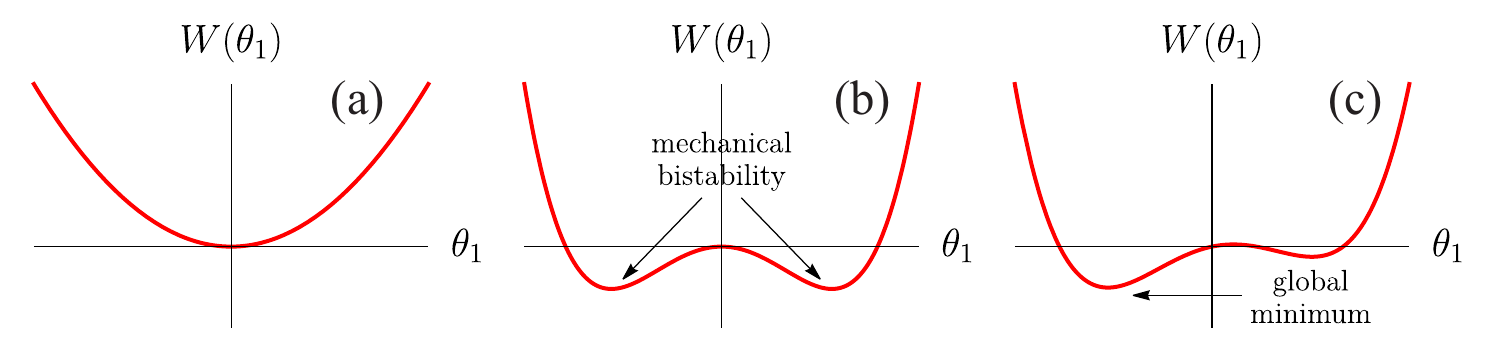}}
\caption{\label{Fig2}%
 The effective energy for the clean nanotube at $\alpha{=}0$  below (a) and above (b) the buckling instability threshold. Above the threshold the instability leads to a mechanical bistability in the clean nanotube.  Disorder breaks the symmetry between two bistable states and the global  energy minimum appears (c), which depends on a specific disorder realization.   }
\end{figure*}

\section{Current for a given curvature \label{Sec:fixed}}

 In order to calculate $\theta_1$ for fixed value of $C_1$ one needs to find all  solutions of the balance  Eq.~\eqref{eq:BA:1}, calculate corresponding energies with the use of Eq.~\eqref{W-theta1} and find global energy minimum by choosing solution with the lowest energy.

 \subsection{Clean case in the absence of electro-mechanical coupling ($C_1{=}0,\alpha{=}0$)}
 
In order to set notations we start from the analysis of a clean nanotube  without electromechanical coupling. 
Then, Eq.~\eqref{eq:BA:1} simplifies drastically,
\be
\theta_1^3-\epsilon \theta_1=0. 
\label{f}
\ee
This equation describes conventional buckling instability~\cite{Landau1970}. It   has a single  stable solution, $\theta_1{=}0$ for $\epsilon{<}0$,
and two stable solutions and an unstable one for  $\epsilon{>}0$: 
\be
  \theta_1= 
  \begin{cases}
       \pm \sqrt{\epsilon},\quad & \text{ (stable)} ,
       \\
            0 ,\quad & \text{ (unstable)}  .
  \end{cases}  
\ee
      The effective energy for $\theta_1$ mode looks:
\be
W(\theta_1)=  \frac{(\theta_1^2-\epsilon)^2}{4}.
\label{W}
\ee

Function $W(\theta_1)$ is plotted  in Fig.~\ref{Fig2}a for $\epsilon{<}0$ and in  Fig.~\ref{Fig2}b for $\epsilon{>}0$. As one can see, the system shows mechanical bistability above the instability threshold, $\epsilon{>}0$.

\subsection{Disordered  case in the absence of electro-mechanical coupling ($C_1 {\neq} 0,\alpha{=}0$)}

Disorder modifies the balance equation and the  effective energy as follows
\begin{align} 
&f(\theta_1)=\theta_1^3-\epsilon \theta_1-8 C_1=0,
\label{theta1-C1}
\\
&
W(\theta_1)=   \frac{(\theta_1^2-\epsilon)^2}{4}-8 C_1 \theta_1.
\label{WC}
\end{align} 
The effective energy for $C_1\neq 0$ is shown in Fig.~\ref{Fig2}c.  Solutions of  Eq.~\eqref{theta1-C1} depend on $C_1,$  so that for $\epsilon{>}0$ and  at sufficiently weak disorder there are two stable solutions, $\theta^\pm(C_1)$ (see blue thick curves in Fig. \ref{fig:3}), and one unstable solution (see dashed curve in Fig. \ref{fig:3}),   just as in the clean case.
It is easy to check that $W(\theta^+){<}W(\theta^-)$  for $C_1{>}0$ and  $W(\theta^+){>}W(\theta^-)$  for $C_1{<}0$.  
The dependence  $\theta_1(C_1)$ corresponding to the absolute energy  minimum of $W(\theta_1)$
is shown in Fig. \ref{fig:3} by red thick curve for $\epsilon>0.$ 
As seen, there is a 
jump at $C_1=0,$ corresponding to  transition from $\theta^-$ to $\theta^+.$ 
Hence, in the case $\epsilon>0$ the dependence of $\theta_1$ on $C_1$ can be written as
\be 
\theta_1(C_1)\!=\! \Theta(C_1) \theta^+(C_1)\!+\!\Theta(-C_1) \theta^-(C_1) .
\label{theta1-C1-abs}
\ee

\subsection{Effect of electro-mechanical coupling}

For $\alpha {\neq} 0,$  the step functions entering  both $f(\theta_1)$ and $W(\theta_1)$  depend on $v_\pm$ [see Eqs.~\eqref{f-theta}, \eqref{nd-theta}, \eqref{W-theta1}, and \eqref{wd}].  As follows from  Eqs.~\eqref{nd-theta} and \eqref{wd}, the quantities $n_{\rm d}$ and $w_{\rm d}$ can take  different  values   depending on relation between $\theta_1$ and $v_\pm$:
\be
 \begin{array}{ll}\displaystyle
     &  n_{\rm d}=1, \quad w_{\rm d}=\theta_1-v_g,   \quad{\rm for}\quad \theta_1<v_-,        \\      &n_{\rm d}=\frac{1}{2},
     \quad w_{\rm d}=\frac{\theta_1- v_+}{2}, \quad \hspace{1mm} {\rm for}\quad v_-\leqslant \theta_1<v_+,
     \\ 
     &n_{\rm d}=0, \quad w_{\rm d}=0, \hspace{1.3cm}{\rm for}\quad v_+\leqslant \theta_1 .
\end{array}
\label{wd-nd}
\ee

\begin{figure}[t]
\centerline{\includegraphics[width=0.7\columnwidth]{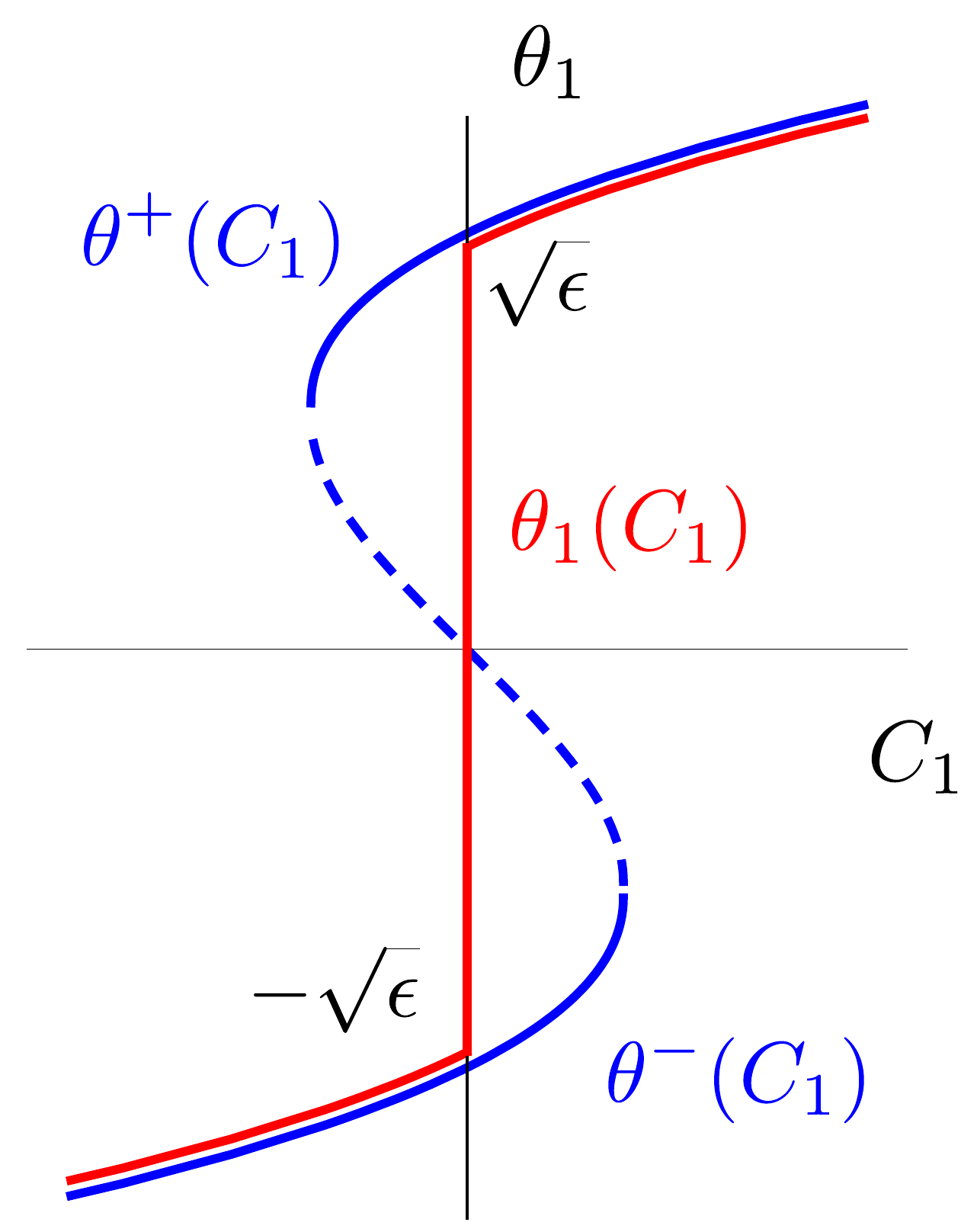}}
\caption{\label{fig:3}
 Schematic dependence of $\theta_1$ on the disorder strength $C_1$, corresponding to the absolute minimum of the effective  energy $W$ (red curve) for $\epsilon>0.$
Blue thick curves represent two stable solutions, $\theta^\pm(C_1), $  dashed curve is the unstable solution. 
}
\end{figure}

\begin{figure}[t]
\includegraphics[width=\columnwidth]
{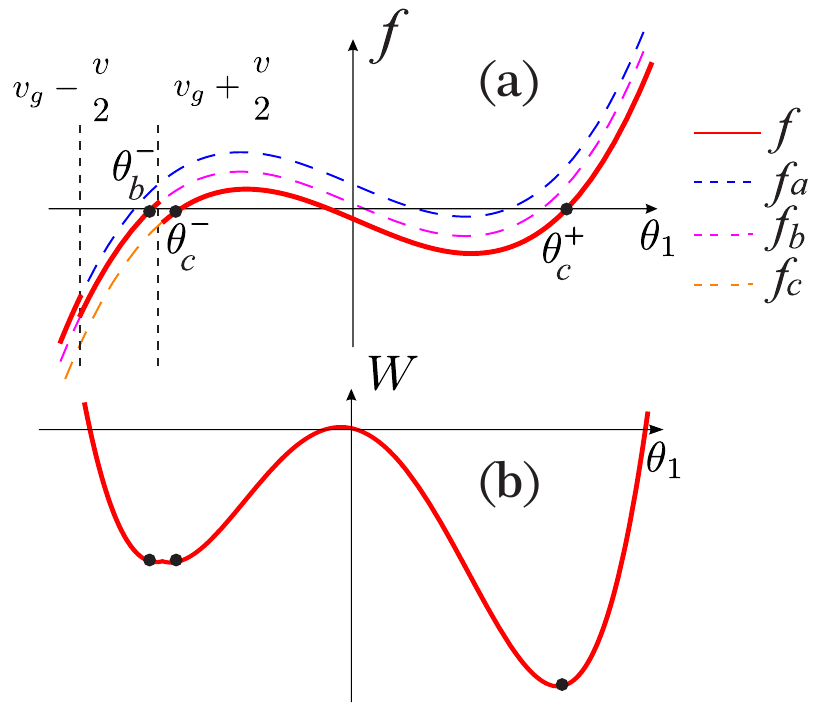}
\caption{\label{f-nd-new}
(a) Graphical  representation of the balance equation
in the presence of electro-mechanical coupling. Dashed lines represent functions $f_a,f_b$ and $f_c$, cf. Eq.~\eqref{f-abc};  
red thick line shows function $f(\theta_1)$, cf. Eq.~\eqref{f-theta}]. Equation $f(\theta_1)=0$  have three solutions: $\theta_b^-,\theta_c^-$ and $\theta_c^+.$ Only one of these solutions, $\theta_b^-,$ corresponds to non-zero current through the nanotube. (b) Energy of nanotube for the 
same voltages.  There are three local minima at  $\theta_b^-,\theta_c^-$ and $\theta_c^+.$  The global minimum correspond to $\theta_c^+,$ so  for such $v_g$ and $v,$ the  current is blocked due to the elastic blockade. }
\end{figure}

It is convenient to introduce three functions,  
\be 
f_s (\theta_1)= \theta_1^3-\epsilon \theta_1 - 8 C_s,  \quad  s=a,b,c,
\label{f-abc}
\ee
that correspond to different possible values of $n_{\rm d}.$ Here, we denote 
\be C_a=C_1-\frac{\alpha}{8},\qquad C_b=C_1-\frac{\alpha}{16},  \qquad C_c=C_1 .
\label{Cabc}
\ee
These functions are shown  in Fig.~\ref{f-nd-new} by dashed lines. As seen from  this figure, 
solutions of equations $f_s(\theta_1){=}0$  give (for not too large magnitudes of $C_1$ and $\alpha$) six values of $\theta_1$ (we do not consider here three unstable states with $\theta_1$ close to zero): \be \theta^\pm_s=\theta^{\pm} (C_s), \label{theta-solutions}\ee
where functions $\theta^\pm(C_1)$ are shown in Fig.~\ref{fig:3}.
We notice that  only two solutions, $\theta^\pm_b,$ belong  to the interval $v_-{<}\theta_1{<}v_+$ and, therefore,   correspond to nonzero current through the nanotube [see Eq.~\eqref{J-theta}].  Using  Eqs.~\eqref{W-theta1} and \eqref{wd-nd},  one can easily find energies corresponding to the solutions $\theta_s^\pm$:  
\begin{gather}
W_a^\pm{=}\omega_a^\pm{-} \alpha v_g,\quad W_b^\pm{=}\omega_b^\pm{-} \alpha v_+/2, \quad W_c^\pm{=}\omega_c^\pm,
\label{W-large-abc}
\\
\omega_s^\pm{=}\frac{[(\theta_s^\pm)^2 {-}\epsilon]^2}{4} {-} 8 C_s \theta_s^\pm.
\label{Wabc}
\end{gather}
Importantly, $\omega_s^\pm$  are functions of $C_1$ and  $\alpha$ only and do not depend on $v_g$ and $v.$

Next step is to find regions in the plane $(v_g,v),$ where one or several solutions \eqref{theta-solutions} are realized and, in the  case of several solutions, chose solution with the lowest energy.   This procedure is illustrated in Fig.~\ref{f-nd-new}.    
Function $f(\theta_1)$ defined by Eq.~\eqref{f-theta} is non-monotonous (see red curve in Fig.~\ref{f-nd-new}a)  and  have jumps at the points $\theta_1{=}v_+$ and $\theta_1{=}v_-.$ 
Equation $f(\theta_1){=}0$ yields several solutions for $\theta_1$ belonging to  manifold \eqref{theta-solutions}. For example,   in Fig.~\ref{f-nd-new}a the 
voltages $v_\pm$ are such that  
there are three solutions: $\theta_b^-,\theta_c^-$ and $\theta_c^+.$
These solutions give the three local 
minima of the energy $W$ (see  Fig.~\ref{f-nd-new}b). As one can see,  the global minimum correspond to   $\theta_c^+,$ so that current through nanotube is zero.  This example illustrates 
 physical origin of the  elastic blockade:   
local minimum $\theta_b^-,$ corresponding to non-zero current, does not give absolute minimum. Hence, an electron placed in this local minimum should decrease its elastic energy   by re-buckling from $\theta_b^-{\approx} {-}\sqrt \epsilon$ to  $\theta_c^+ {\approx} \sqrt \epsilon$   with  simultaneous   "jumping out" from the current-carrying window.

\begin{figure}[t]
\includegraphics[width=\columnwidth]
{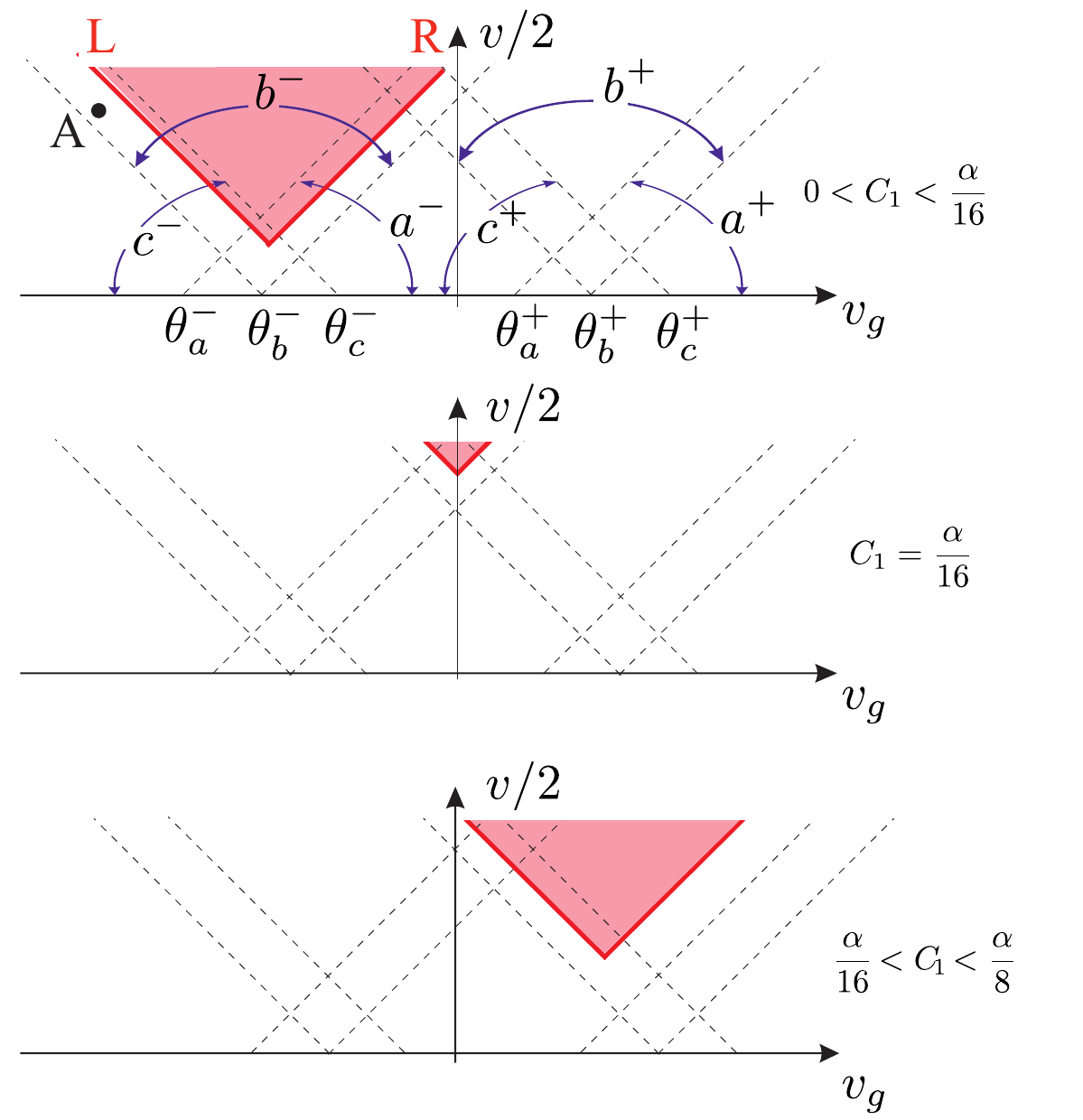}
\caption{\label{phase-diagram}
  Phase diagram with  schematically shown regions of different solutions $\theta_s^\pm.$ As seen, these regions overlap. In the area of overlapping, several solutions can coexist, giving  several local minima of energy (see also Fig.~\ref{f-nd-new}b).  Pink 
  triangle shows the region with non-zero current.  Thick red lines encode dependence of the 
 threshold voltage $v_{\rm Th}$  on $v_g.$ At these lines current ``jumps'' from $I{=}0$ (for $v{<}v_{\rm Th}$) to $I{=}1/2$ (for $v{>}v_{\rm Th}$).  The  evolution of current-carrying 
 triangle with increasing  of $C_1$ is illustrated in top, central, and bottom panels.  For  $0{<}C_1{<}\alpha/16$ the 
 triangle moves up (top panel), reaches maximal position at $C_1{=}\alpha/16$ (central panel), and moves down with further increase of disorder (bottom panel). }
\end{figure}

Let us now discuss the general case. First we notice, that   $(v_g,v)$ plane can be divided onto several regions, where different solutions \eqref{theta-solutions} can be realized.  These regions corresponds to different regions in Eq.~\eqref{wd-nd}. For example, solution $\theta_a^+$ exists if $\theta_a^+{<}v_-,$  solution $\theta_b^+$ if $ v_-{<}\theta_b^+{<}v_+,$ etc. Since $\theta_s^\pm$ do not depend on $v_g$ and $v,$ these regions are limited by lines: $v/2{=}\pm v_g {+} {\rm const}.$  They are shown by dashed lines  in Fig.~\ref{phase-diagram}, where  we plotted only  half plane $v{>}0$ (the diagram is symmetric  with respect to change $v {\to} -v$  and change  $I {\to} -I$). Dashed lines connected by arrowed arcs 
indicates regions where different solutions \eqref{theta-solutions} exist, e.g. regions $a^\pm$ corresponds to solutions $\theta^\pm_a$ etc.
As seen, these regions overlap, so that several solutions can coexist in agreement with Fig.~\ref{f-nd-new}. For example, in the points $A$  three solutions coexist, $(c^-,b^-,c^+).$ 
Therefore,  one has to calculate energies of coexisting states ($W_c^-,W_b^-,$ and $W_c^+$ in the above example) and to find the global minimum. Within the pink triangle  in Fig.~\ref{phase-diagram} 
the global minimum is given by 
one of the energies $W_b^-$ or $W_b^+,$ corresponding to non-zero current, $I=1/2.$ 
Boundary of this 
triangular region, shown by two red thick lines, represents the threshold voltage, $v_{\rm Th}{=}v_{\rm Th}(C_1,v_g).$ This voltage separates the region with zero 
current ($v{<}v_{\rm Th}$)  from the region ($v{>}v_{\rm Th}$),  where $I=1/2.$ Position of the  
triangle depends on $C_1$ in a non-trivial way.  Analyzing Eqs.~\eqref{W-large-abc} and \eqref{Wabc} one can demonstrate that the left (L) and the right (R) boundaries (red lines) are determined by the following conditions: 
\be
\begin{aligned}
&L:W_b^-{=}W_c^-,\quad  R:W_b^-{=}W_a^-, \quad {\rm for}~ C_1{<}0,     
\\
&L:W_b^-{=}W_c^+,\quad  R:W_b^-{=}W_a^-, \quad{\rm for}~ 0{\leqslant}C_1{<}{\alpha}/{16},
\\
&L:W_b^+{=}W_c^+,\quad  R:W_b^+{=}W_a^-, \quad{\rm for}~{\alpha}/{16}{\leqslant}C_1{<}{\alpha}/{8},
\\
&L:W_b^+{=}W_c^+,\quad  R:W_b^+{=}W_a^+, \quad{\rm for}~ {\alpha}/{8}{\leqslant}C_1.
\end{aligned}
\label{LR-Wabc}
\ee
Using these relations and Eq.~\eqref{W-large-abc}, we find that the threshold voltage can be written as follows:
\be
v_{\rm  Th}=v_0(C_1) + 2|v_g -v_{g0}(C_1)|. 
\label{vth-C1}
\ee
Dependence of  $v_{\rm Th}{=}v_{\rm Th} (v_g,C_1)$ on $v_g$ for fixed $C_1$ is shown by red lines (boundaries of the current-carrying triangle) 
in Fig.~\ref{phase-diagram}.
 The dependence on $C_1$ is  fully encoded in functions   $v_0{=}v_0(C_1)$ and $v_{g0}{=}v_{g0}(C_1)$:
\be
v_0=\frac{2}{\alpha}\left\{ 
\begin{array}{ll}
     &2\omega_b^- {-}\omega_c^-{-}\omega_a^-  , \quad{\rm for}~ C_1{<}0,
     \\
     &2\omega_b^- {-}\omega_c^+{-}\omega_a^-, \quad{\rm for}~ 0{\leqslant}C_1{<}{\alpha}/{16},
     \\
     & 2\omega_b^+ {-}\omega_c^+{-}\omega_a^-, \quad{\rm for}~{\alpha}/{16}{\leqslant}C_1{<}{\alpha}/{8},
     \\
     &2\omega_b^+ {-}\omega_c^+{-}\omega_a^+, \quad{\rm for}~ {\alpha}/{8}{\leqslant}C_1,
     \end{array}
\right.
\label{v0}
\ee
\be
v_{g0}=\frac{1}{\alpha}\left\{ 
\begin{array}{ll}
     &\omega_a^- -\omega_c^- , \quad{\rm for}~ C_1{<}0,
     \\
     &\omega_a^- -\omega_c^+, \quad{\rm for}~ 0{\leqslant}C_1{<}{\alpha}/{8},
     \\
          &\omega_a^+ -\omega_c^+, \quad{\rm for}~ {\alpha}/{8}{\leqslant}C_1,
     \end{array}
\right.
\label{vg0}
\ee

Equations \eqref{vth-C1}, \eqref{v0}, and \eqref{vg0} provide a full solution of the problem for a fixed disorder. Functions $\omega_s^\pm$  entering these equations are given by Eq.~\eqref{Wabc} with   $\theta^\pm_{s}$ found from   
Eq.~\eqref{theta-solutions}. 

For a fixed disorder the current-voltage  
dependence becomes
\be
I(v)=\frac{{\rm sign} (v)}{2} \Theta (|v| -v_{\rm Th}).
\label{I-V-C}
\ee

Equations ~\eqref{v0} and \eqref{vg0}  allows one to find some general properties
of  the  current-carrying triangle.  
In particular, one can easily demonstrate 
that the curve $v_{0}(C_1)$ has a maximum at $C_1{=}\alpha/16$ and  is symmetric with respect to this point: $v_{0}(C_1){=}v_{0}(\alpha/8{-}C_1).$  Also, one finds $v_{ g0}{=}0$ at $C_1{=}\alpha/16$ and is anti-symmetric with respect to this point
$v_{g0}(C_1){=}{-}v_{ g0}(\alpha/8{-}C_1).$ 

  \begin{figure}[t]
\includegraphics[width=0.83\columnwidth]{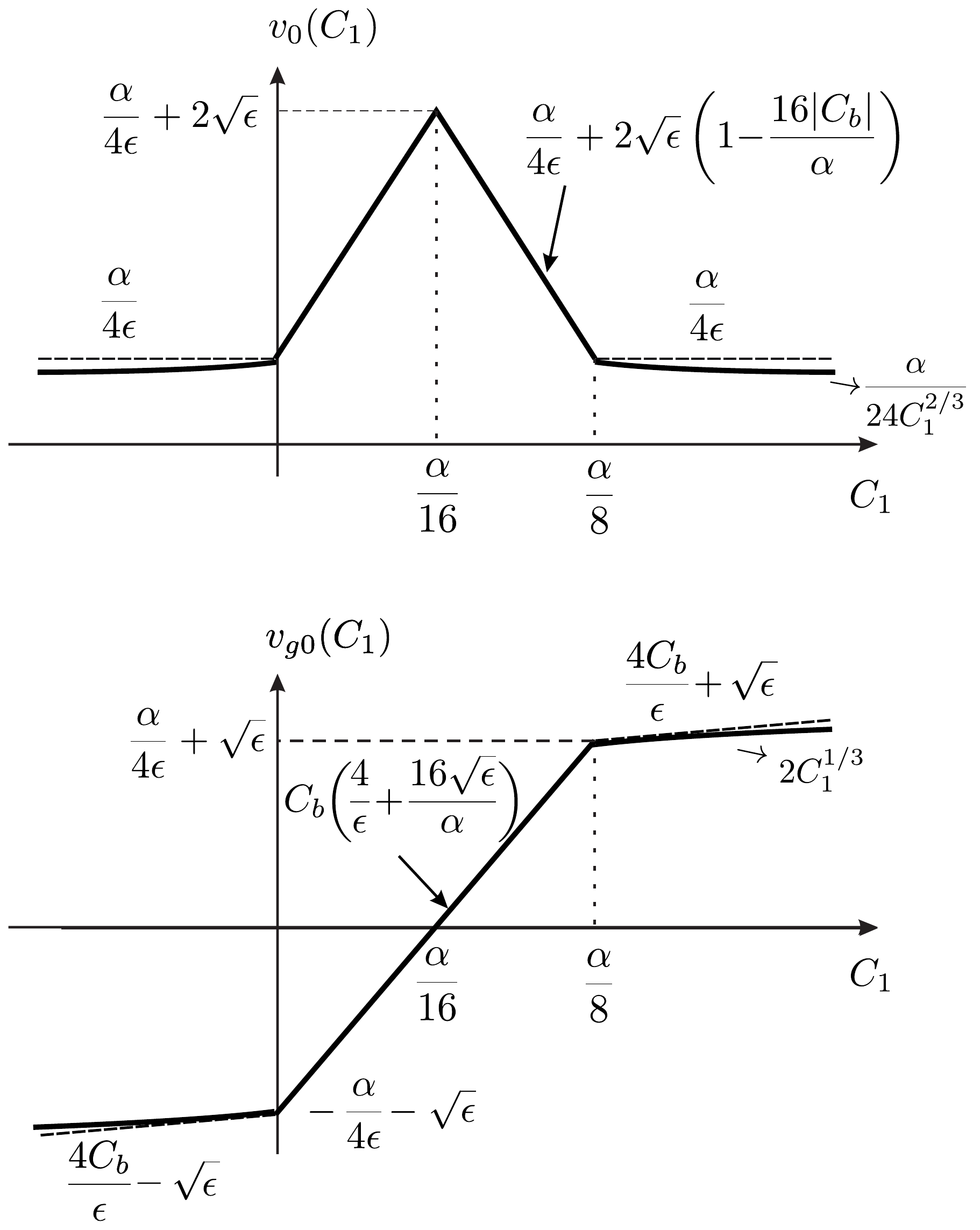}
\caption{\label{v0vg0}
Dependence of position  of the lower tip of the current-carrying 
triangle
on the curvature.  Black thick lines are dependencies of $v_0$ (top panel)  
and $v_{g0}$ (bottom panel) on  $C_1$ for weak disorder and weak electro-mechanical coupling $\alpha$ ($C_1{\ll} \epsilon^{3/2},$  $\alpha {\ll} \epsilon^{3/2}$).  At larger $C_1$  ($|C_1| {\gg} \epsilon^{3/2} $), $v_0$ starts to decrease very slowly, ${\sim} \alpha/C_1^{2/3},$   while linear increase of $v_{g0}$ with $C_1$ slows down to ${\sim} C_1^{1/3}.$ Dashed lines show asymptotical behavior of $v_0$ and  $v_{g0}$  for  $C_1\ll \epsilon^{3/2} .$  Thick curves deviate significantly from dashed ones for $|C_1| \sim \epsilon^{3/2} \gg \alpha.$   }
\end{figure}

\subsection{Limiting cases}

Next, we consider some limiting cases.   General equations derived in the previous section are dramatically simplified when  $C_1$ and $\alpha$ are small:
  \be
  C_1 \ll \epsilon^{3/2},\quad \alpha \ll \epsilon ^{3/2}. 
 \label{small-C1-alpha}
  \ee
  Then, keeping  in  $\theta^\pm_{s}$    terms  up to linear order with respect to $C_1$ and $\alpha,$    and in  $\omega_s^\pm$  up to  quadratic order   we get
  \begin{align}
      & \theta_s^\pm \approx \pm \sqrt \epsilon +\frac{4 C_s}{\epsilon},\label{theta-s}
      \\
      & \omega_s^\pm \approx-\frac{16 C_s^2}{\epsilon} \mp 8 C_s \sqrt \epsilon .\label{omega-s}
  \end{align}
  Substituting Eq.~\eqref{omega-s} into Eqs.~\eqref{v0} and \eqref{vg0} we find that the latter equations can be written in a compact way (it is convenient to express $v_0$ and $v_{g0}$ in terms of $C_b$:
    \be
v_0(C_1){=}
\begin{cases}
\displaystyle
\frac{\alpha}{4\epsilon}, & \displaystyle \text{for}~|C_b|{>}\frac{\alpha}{16}\\
\displaystyle \frac{\alpha}{4\epsilon} +2 \sqrt \epsilon\left( 1{-}\frac{16|C_b|}{\alpha}\right), &\displaystyle \text{for}~|C_b|{<}\frac{\alpha}{16} ,
\end{cases}
\label{v0-small-C1}
\ee
  \be
v_{g0}(C_1){=}
\begin{cases}
\displaystyle
\frac{4  C_b}{\epsilon}{+} {\rm sign}( C_b)\sqrt \epsilon, & \displaystyle \text{for}~|C_b|{>}\frac{\alpha}{16}\\
\displaystyle C_b \left(\frac{4}{\epsilon}{ +} \frac{16 \sqrt \epsilon  }{\alpha} \right), &\displaystyle \text{for}~|C_b|{<}\frac{\alpha}{16} .
\end{cases}
\label{vg0-small-C1}
\ee

Dependencies of  $v_0$ and $v_{g0}$ on $C_1$ are schematically shown in Fig.~\ref{v0vg0}.
Most interestingly, there is a very sharp increase of $v_0$ (and, consequently, $v_{\rm Th}$) within a narrow region of disorder strength: $0{<}C_1{<}\alpha/8.$
Hence, we predict the suppression of the current  by the built-in curvature--- the phenomenon, which we call \emph{elastic curvature blockade}. Physically, curvature-induced  
threshold in the current arises due to the need to choose the bending amplitude 
$\theta_1$  from 
minimization of $W(\theta_1)$. 
For $|v|{<}v_{\rm Th}$, the bending angle found from  energy minimization  (with account of the electro-mechanical coupling) turns out to be beyond the current-carrying window $v_- {<}\theta_1{<}v_+.$

\begin{figure}[t]
\includegraphics[width=0.94\columnwidth]{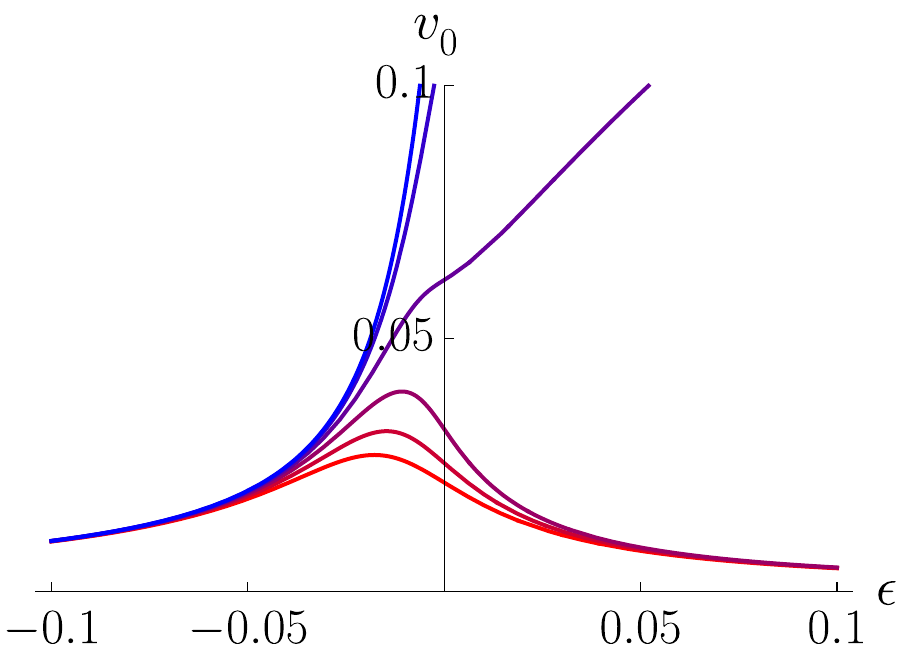}
\caption{\label{fig:vTh}
Dependence of the   elastic-blockade-induced  voltage, $v_{0}$, on $\epsilon$  for  $\alpha{=} 2\cdot10^{-3}$  at different values of $C_1$ ranging from $-1.25\cdot 10^{-4}$ to $1.25\cdot 10^{-4}$ (from red to blue) with the step $0.5\cdot 10^{-4}$.   For $\epsilon{>}0,$ the built-in curvature  enhances parametrically the  voltage $v_0.$ }
\end{figure}

One can also find dependence of $v_0$ on $\epsilon.$  The results of corresponding numerical 
solution is shown in Fig.~\ref{fig:vTh}.   As one can see, this dependence is very sensitive to
a built-in curvature. In particular, 
the change of $C_1$ from  negative to positive values   results in qualitative change of the behavior of $v_{\rm 0}$ with $\epsilon$.

  Equations \eqref{v0-small-C1} and \eqref{vg0-small-C1} can be further simplify in the absence of disorder, $C_1{=}0.$ In this case, $v_0{=}\alpha/4 \epsilon$ $v_{g0}{=}{-}\sqrt \epsilon {-}\alpha/4\epsilon ,$  so we obtain 
  \be
  v_{\rm Th}=\frac{\alpha}{4 \epsilon} +
  2\left|v_g + \sqrt \epsilon + \frac{\alpha}{4 \epsilon}\right|, \quad{\rm for}~C_1=0.
  \label{Vth-C=0}
  \ee
 This result is in agreement with Refs.~\cite{Weick2010, Weick2011,Micchi2015}, where it was found that    electro-mechanical coupling  leads to elastic blockade.  Indeed, we see that there is a finite  minimal threshold voltage, 
 $v_{\rm Th}^{\rm min}={\alpha}/{4 \epsilon},$ proportional to electro-mechanical coupling.  This means  that at low temperatures  and small  driving voltage, the    transport through   SET is blocked.    
 
 In the  limit  of very small $\alpha,$ one can easily get $I-V$ curve for arbitrary values of $C_1.$ To this end, one should take limit $\alpha{\to}0$ in Eqs.~\eqref{v0}, and \eqref{vg0}. Simple calculation yields
 $v_0{\to}0$ and $v_{g0}{\to}\theta_1(C_1)$ where  $\theta_1(C_1)$ is given by  Eq.~\eqref{theta1-C1}. Therefore, we get
 \be
 v_{\rm Th}=2 \left|v_g -\theta_1(C_1)  \right|, \quad{\rm for} \quad \alpha \to 0 .  \ee
  Since $\theta_1(C_1)$ 
  has a jump at $C_1{=}0$ (see Fig.~\ref{fig:3}), current-carrying triangle  
  also abruptly  ``jumps'' in $v_g$ 
direction on $\delta v_g{=}2 \sqrt \epsilon,$ when $C_1$ changes sign.
Exactly at the point $v_g{=}0$  and for $|C_1|{\ll}\epsilon^{3/2},$ we get $v_{\rm Th}{\approx}2 \sqrt \epsilon.$
 Comparing  this value  with the value $ \alpha/4\epsilon$  for  finite $\alpha$ and $C_1{\equiv}0$ [see Eq.~\eqref{Vth-C=0}], we find giant, about $ \epsilon^{3/2} \alpha^{-1}{\gg}1,$ 
curvature-induced enhancement of the driving voltage threshold, as compared to  the elastic blockade threshold, in the absence of the curvature  
\cite{Weick2010,Weick2011,Micchi2015}. 

Before closing this section, we note that   Eqs.~\eqref{v0-small-C1} and \eqref{vg0-small-C1}  have finite limit when both    $C_1$ and $\alpha$ tends to zero, while the ratio $C_1/\alpha{=}(1+\xi)/16$ remains finite. Then, we find
\be
\begin{aligned}
&\frac{v_0}{2 \sqrt \epsilon}=\left\{\begin{array}{ll}
      1-|\xi|,&\quad{\rm for}~ |\xi|<1, \\
      0,&\quad{\rm for}~ |\xi|>1,
\end{array} \right.
\\
&\frac{v_{g0}}{ \sqrt \epsilon}=\left\{\begin{array}{ll}
      \xi,&\quad{\rm for}~ |\xi|<1, \\
      {\rm sign}[ \xi],&\quad{\rm for}~ |\xi|>1.
\end{array} \right .
\end{aligned}
\label{xi-approximation}
\ee
   Physically, approximation \eqref{xi-approximation} neglects relatively small effects caused  by elastic blockade in the absence  of curvature, but  captures  much larger curvature-induced  blockade. Within this approximation,  position of the current-carrying triangle
      is fully determined by parameter   $\xi= 16 C_{\rm b}/\alpha{=}16 C_{\rm 1}/\alpha{-}1.$   With increasing $ \xi$ from ${-}\infty$ to $\infty$ the lower tip of the triangle 
      moves in the following way:   it remains stationary in the position $(v_{g0}{=}{-}\sqrt \epsilon, v_0{=}0)$  for $ {-}\infty{<}\xi{<}{-}1,$ then  ``climbs a hill'' with a height $2\sqrt \epsilon$ in the interval ${-}1 {<}\xi{<}0,$  moves down   in the interval $ 0{<}\xi{<}1,$  and finally    arrives at the position $(v_{g0}{=}\sqrt \epsilon, v_0{=}0)$ and remains stationary for   $ 1 {<}\xi{<}\infty.$ 
   This approximation neglects very slow motion of the triangle 
    at large $|\xi|{>}1$ (see the bottom panel in Fig.~\ref{v0vg0}).    The change of $v_{g0}$ from ${-}\sqrt \epsilon$ to                    $\sqrt \epsilon$ with increasing $\xi$ is due to curvature-induced re-buckling of the nanotube.

\section{Disorder-averaged $I-V$ curve\label{Sec:random}}  
 
\begin{figure}[h!]
	\includegraphics[width= \linewidth]{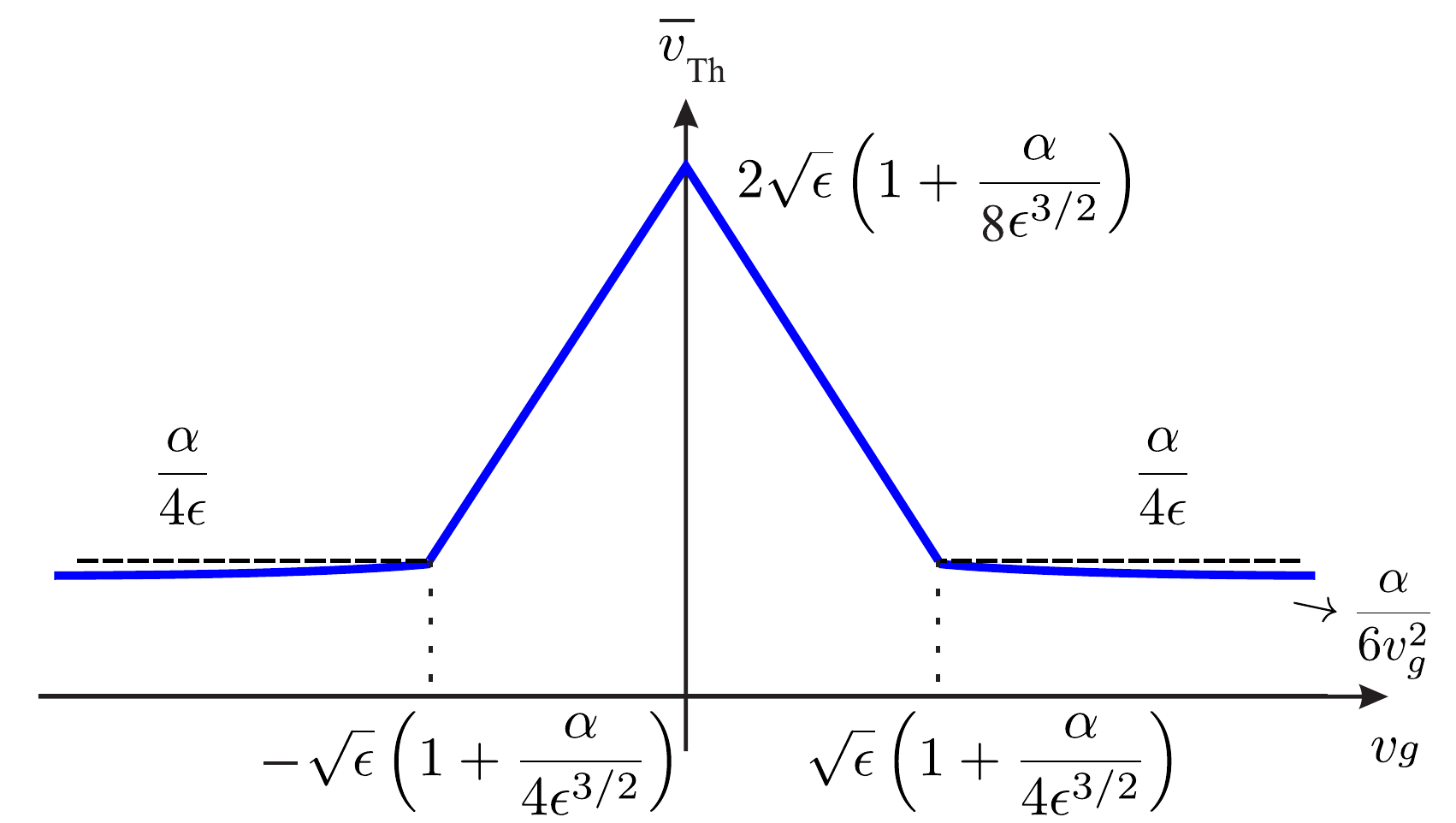}
	\caption{Dependence of the threshold voltage,   $\overline{v}_{\rm Th} $ on $v_g.$ 
	Below the curve $\overline{v}_{\rm Th}(v_g)$ disorder-averaged current equals to zero. For $0<|v_g| -\sqrt \epsilon \ll \sqrt \epsilon,$  the thershold voltage is approximately constant and given by $\alpha/4\epsilon$ (dashed line).   }
\label{vth}
\end{figure}

 Next we assume that  a random $C_1$ 
  is described by a distribution function $\mathcal P(C_1).$     
 Changing  $C_1$ in the interval, ${-}\infty {<}C_1{<}\infty,$  one can find the curve $(v_g(C_1),v_0(C_1))$ in $(v_g,v)$  plane parameterized  
by $C_1.$ This curve yields limiting value $\overline{v
}_{\rm Th} (v_g),$ below which the current  is equal to
zero for any $C_1$ and consequently, for disorder-averaged $I-V$ curve.
Using general properties of  functions  $v_0(C_1)$ and $v_{g0}(C_1)$ 
[see discussion after Eq.~\eqref{I-V-C}],  one can easily find that  $\overline{v
}_{\rm Th} (v_g)$ has a maximum at $v_g{=}0$ and is symmetric with respect 
to this point:   $\overline{v
}_{\rm Th} (v_g){=}\overline{v
}_{\rm Th} ({-}v_g).$ Dependence $\overline{v
}_{\rm Th} (v_g)$ is  shown schematically  in Fig.~\ref{vth}.   
For small $\alpha$  obeying inequality  \eqref{small-C1-alpha},  this dependence  reads
\begin{equation}
    \overline{v}_{\rm Th}(v_g)=
    \begin{cases}
    \displaystyle \frac{\alpha}{4 \epsilon}, & \text{for}~|v_g|{>}\overline{v}_g , \\
    \displaystyle 2\sqrt \epsilon \left (1{+} \frac{\alpha}{8 \epsilon ^{3/2}}\right )
                     {-} \frac{2 |v_g|}{1{+} \frac{\alpha}{4 \epsilon ^{3/2}}},  & \text{for}~|v_g|{\leqslant}\overline{v}_g  ,
    \end{cases}
    \label{vth-vg}
\end{equation}
where $\overline{v}_g{=}\sqrt \epsilon [1{+} \alpha/(4 \epsilon^{3/2})]$.
 One can also show that for very large gate voltages,  $v_g {\gg} \sqrt\epsilon,$ the  threshold voltage  decays as $\overline{v}_{\rm Th}(v_g) {\approx} \alpha/(6v_g^2).$  

We stress that $\overline{v}_{\rm Th}(v_g)$   does not depend on the distribution function of the random curvature, $\mathcal P(C_1),$ provided that this function  is non-zero within  the whole interval $-\infty{<}C_1{<}\infty.$   
To be specific, 
   we assume that the random curvature has Gaussian distribution with the zero mean and 
\begin{equation}
\langle C^\prime(s_1)C^\prime(s_2)\rangle = \Delta \delta(s_1-s_2),   
\label{eq:CC:cor}
\end{equation}
where $C'(s){=}dC(s)/ds.$
 Then the eigenmodes $C_n$ are independently correlated with the dispersion determined by $\Delta$, 
 \be \label{corr-Cn}
    \left\langle C_n C_m \right\rangle = \Delta_n \delta_{nm},\quad \Delta_n = \frac { 2 \Delta L } { \pi^2 n^2 },\quad n=1,2,\dots.
\ee
Hence, $C_1$  has the normal distribution, 
\be \mathcal P(C_1) = \frac{\exp\left(-{C_1^2}/{2\Delta_1}\right)}{\sqrt{2\pi \Delta_1}},
\label{P-C1-normal}
\ee 
characterized by the variance  $\Delta_1.$   

The point 
$\{v_{\rm g0}(C_1), v_{0}(C_1)\}$ is the position 
 of the  
 lower tip of triangle  region of non-zero current.  This point  moves as $C_1$ is increasing from negative to positive values.  As shown in  Fig. \ref{fig:1},  for $v{>}\overline{v}_{\rm Th}(v_g),$ there are two values of curvature $C_{\rm R}$ and $C_{\rm L}$ ($C_{\rm L}{>}C_{\rm R}$) for which, respectively, right and left boundaries of the current-carrying triangle 
 cross a point $(v_g,v).$  
  Since at any point belonging to the triangle
  the current equals $I=1/2,$  and equals to zero outside  the triangle,  
  we  get the following expressions for averaged current and the current distribution function 
\begin{gather}
\langle I \rangle {=} 
\frac{1}{2} 
\int\limits_{C_{\rm R}}^{C_{\rm L}} dC_1 \mathcal P(C_1) {=}\frac{1}{2} \left [\erf \left (\frac{C_{\rm L}}{\sqrt{2 \Delta_1}}\right ) {-} \erf\left ( \frac{C_{\rm R}}{\sqrt{2 \Delta_1}}\right )\right],
\notag 
\\
\mathcal{P}_{\rm cur}(I) =  2  \langle I\rangle \delta(I-1/2) +(1- 2  \langle I\rangle )\delta(I) .
\label{eq:av:cur}
\end{gather}
As follows  from Eqs.~\eqref{vth-C1}, 
$C_{\rm R,L}$ are implicitly defined by the following equations:
\be
v_-=v_{g0}(C_{ \rm R})- \frac{v_0(C_{\rm R})}{  2},\quad v_+=v_{g0}(C_{ \rm L})+ \frac{v_0(C_{\rm L})}{  2}.
\label{C-RL}
\ee
\begin{figure}[h!]
\includegraphics[width=0.9 \linewidth]{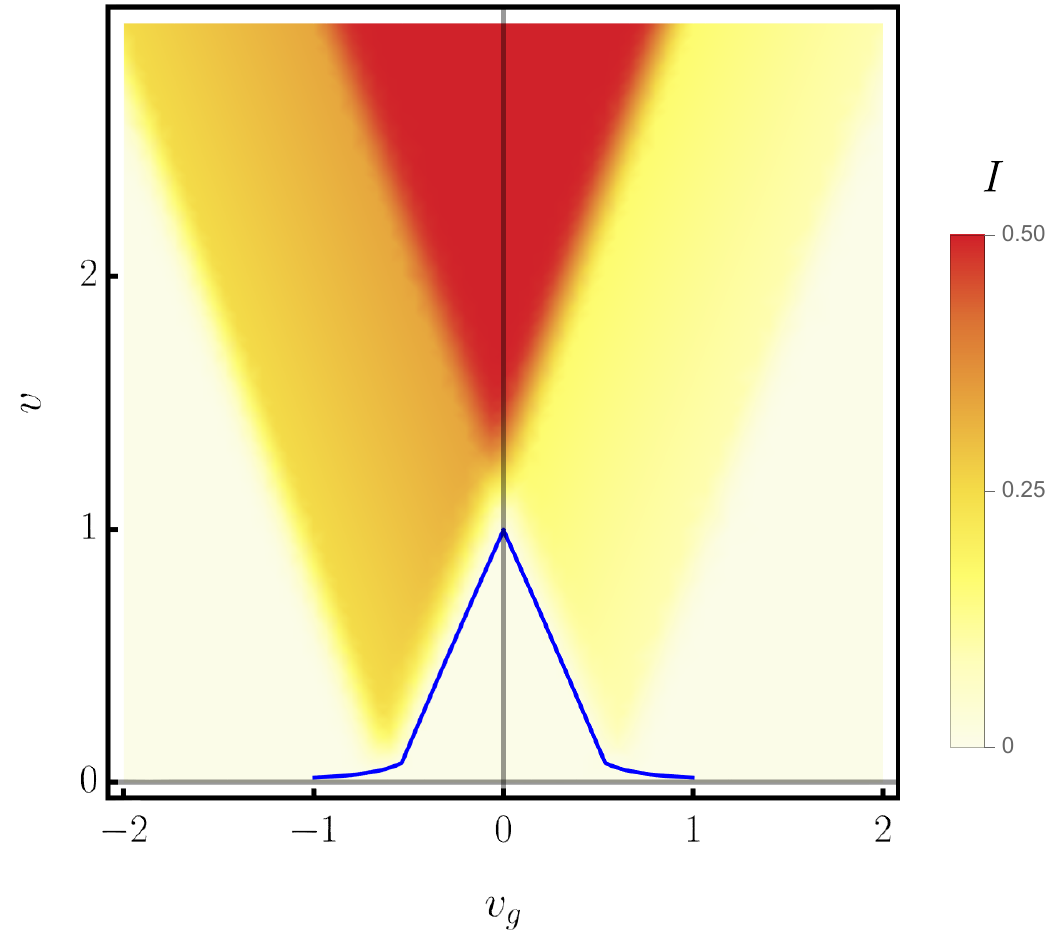}
\includegraphics[width=0.9 \linewidth]{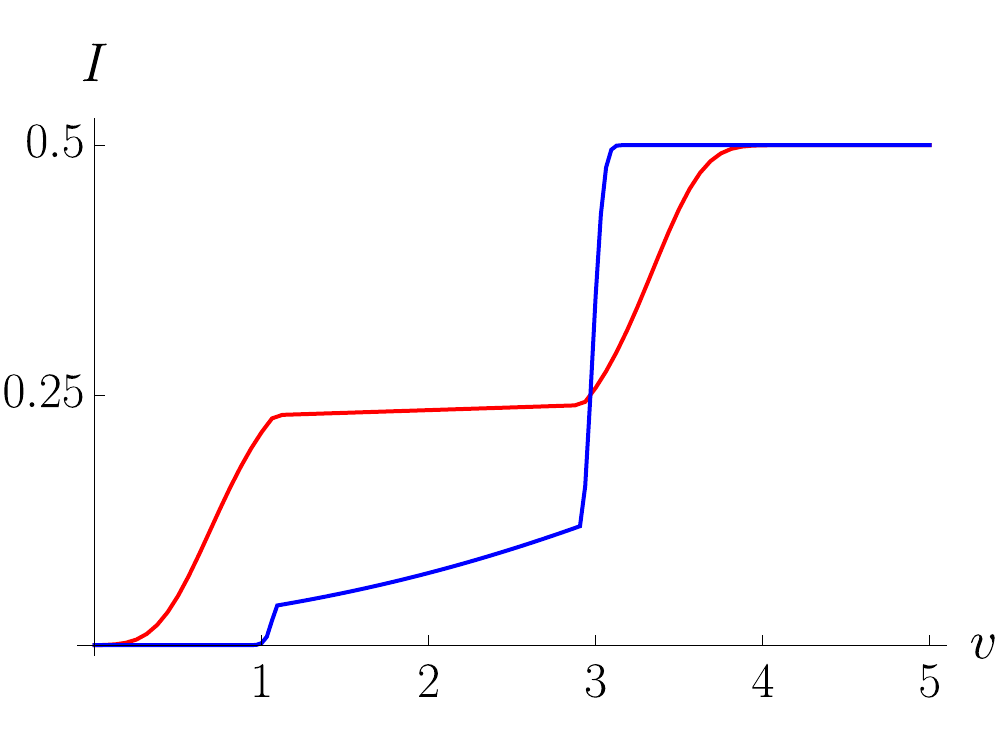}
\caption{
\label{fig:6} 
Top panel: Density plot of the averaged current for $\epsilon{=}0.2$, $\alpha{=}0.1$, and $\Delta_1{=}2\cdot 10^{-4}.$ The blue solid curve is the threshold $\overline{v}_{\rm Th}(v_{\rm g})$.    
Bottom panel: The current-voltage characteristics for $v_{\rm g}{=}1,$
$\alpha{=}0.02$  and  different values of disorder  variance   $\Delta_1{=}8\cdot 10^{-4}$ and $\Delta_1{=}3\cdot 10^{-6}$ (red and blue correspondingly) for zero temperature. As seen, with increasing $\Delta_1$ an intermediate plateau appears corresponding to $I=1/4.$  }
\end{figure}
Here  functions 
$v_0(C)$ and $v_{g0}(C)$ are determined by Eqs.~\eqref{theta-solutions}, \eqref{Wabc}, \eqref{v0}, and \eqref{vg0}.   
Equation~\eqref{eq:av:cur} suggests that 
$\langle I \rangle$ cannot exceed the value $1/2$. 

The density plot of the  average current  calculated numerically with the use of Eq.~\eqref{eq:av:cur} as the function of $v$ and $v_{\rm g}$ is shown in the top panel of Fig.~\ref{fig:6}. Numerically calculated current-voltage curve is presented in the bottom  panel of Fig.~\ref{fig:6} for two different values of variance $\Delta_1.$      One can also obtain  analytical  solution for  $I-V$  by using  approximation \eqref{xi-approximation}.  Corresponding calculations are delegated to Appendix~\ref{simple-model}.    

Analysis of the numerical and analytical solutions demonstrate that  at fixed $v_{\rm g}$, the value $1/2$ for the current is reached at large values of $v$ when  $C_{\rm L} {\to}  \infty,~C_{\rm R} {\to}{-}\infty$.  One of the most interesting properties of  the averaged $I-V$ curve is appearance of an intermediate plateau at $I{=}1/4$ with increasing disorder variance. This property is illustrated in the bottom panel of  Fig.~\ref{fig:6} and in Fig.~\ref{Fig-step-simplified}. In order to understand physics behind this phenomenon,  we will discuss in the next section the distribution function  of the angle $\theta_1.$       

\section{The distribution function of the bending angle, $\theta_1$\label{Sec:theta-PDF}} 
The probability distribution function (PDF) for  the bending angle $\theta_1$ is given by   
\be
 P(\theta_1) = \bigl \langle  \delta[\theta_1-\theta_1(C_1)] \bigr \rangle_{C_1},
\label{P-C1}
\ee
where function $\theta_1(C_1)$ gives solution of  equation $f(\theta_1)=0,$ corresponding to the global minimum of $W$
and angular brackets stand for the  averaging with function $\mathcal P(C_1)$ [see Eq.~\eqref{P-C1-normal}].  Average current can be found (instead of integration over $C_1,$ according to Eq.~\eqref{eq:av:cur}) with the  use  of  $P(\theta_1)$  
by integration 
within the current-carrying window:
\be
\langle I \rangle= \int \limits _{v_-}^{v_+}  P(\theta_1)d\theta_1.
\ee

\subsection{Distribution of the bending angle in the absence  of  electro-mechanical coupling}

Let us first consider the case of zero electro-mechanical coupling. 
For $\alpha{=}0,$ 
 $\theta_1(C_1)$  
is shown in Fig. \ref{fig:3} and for $\epsilon{>}0$  is expressed in terms of $\theta^\pm(C)$ according to Eq.~\eqref{theta1-C1-abs}.

Integration of the delta functions in Eq.~\eqref{P-C1} over $C_1$ yields 
\begin{align}
\label{PthetaC1}
&    P(\theta_1) =    \frac{1}{\sqrt{2\pi \Delta_1}}  \left[\frac {d\theta_1 (C_1) }{dC_1}\right]^{-1}  
    e^{-[C_1(\theta_1)]^2/(2\Delta_1)} 
    \\
    &= \frac{(3 \theta_1^2-\epsilon )~\Theta( \theta_1^2-\epsilon)}{8\sqrt{2\pi \Delta_1}}\exp\left[-\frac{(\theta_1^3-\epsilon \theta_1)^2}{128 \Delta_1}\right] .
\label{P-C3=0}
      \end{align}
Here, we took into account that  $C_1{=}C_1(\theta_1)$ is given by  Eq.~\eqref{theta1-C1} and used the following equation
\be
J(\theta_1) =\frac{dC_1(\theta_1)}{d\theta_1}= \frac{3 \theta_1^2-\epsilon}{8} 
\ee
for 
the   Jacobian (we write here $J(\theta_1)$ instead of $|J(\theta_1)|$, since $J(\theta_1){>}0$ in the region $|\theta_1|{>}\sqrt{\epsilon}$ where $P(\theta_1){>}0$.).  The $\Theta$-function entering  Eq.~\eqref{P-C3=0} reflects  opening a gap in the distribution function.

 In the limit  of a very weak   disorder, the distribution function consists of two delta peaks,
\be
P(\theta_1)\big|_{\Delta_1\to 0}=\frac{1}{2}
     \Bigl [\delta(\theta_1-\Delta_\epsilon)+\delta(\theta_1+\Delta_\epsilon)\Bigr] .  
    \label{P-C=0}
    \ee
Here $\Delta_\epsilon{=} \sqrt \epsilon~\Theta (\epsilon)$
is the  buckling-induced gap. The disorder broadens the delta functions in Eq. \eqref{P-C=0}. 

\begin{figure}[t]
\includegraphics[width=0.95\columnwidth]{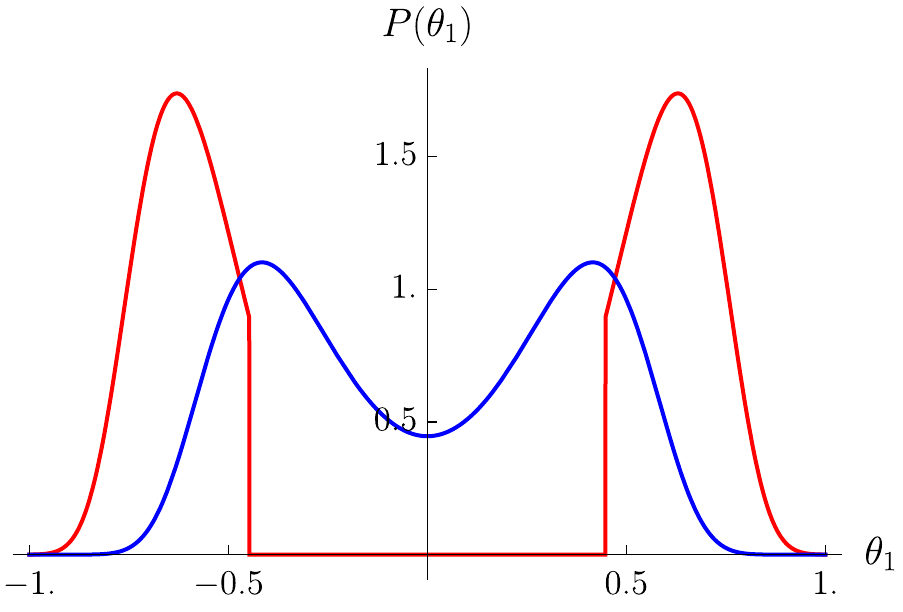}
\caption{
\label{PDF-alpha=0}
 The distribution function for $\alpha{=}0,$ 
as determined by Eq.~\eqref{P-C3=0}, for a fixed weak disorder, $\Delta_1{=}5\cdot 10^{-4}$, and for $\epsilon{=}\pm 0.2$ (red and blue curves, respectively)}
\end{figure}

The function $P(\theta_1)$,  described by Eq.~\eqref{P-C3=0}, is plotted in  Fig.~\ref{PDF-alpha=0}.  for a fixed weak disorder and 
two values of  $\epsilon,$  below and above instability threshold. 
Above the  threshold, $\epsilon{>}0$, there is a gap $|\theta_1|{<}\Delta_\epsilon$ inside which     $P(\theta_1){\equiv}0$. We also notice that slightly  above the gap the distribution function   increases despite of the presence of the damping exponent
in Eq.~\eqref{P-C3=0}. This increase 
is due to the  Jacobian, $J(\theta_1).$

Although the terms involving  higher curvature harmonics 
$C_{n>1}$ are  small, they can  modify the result~\eqref{P-C3=0} in the region where $P(\theta_1)$  is very small. In particular, they  lead to appearance of tails inside the gap, $|\theta_1|{<}\Delta_\epsilon$, for $\epsilon{>}0$ and to some modification of $P(\theta_1)$  for $\epsilon{<}0$. We discuss in detail the effects of the higher harmonics in Appendix \ref{high-harm} and demonstrate that these modifications  do not lead to significant change of the distribution function.  In particular,  as one can see from Fig.~\ref{cross-P-f}, even in the case of a 
moderate disorder, when  delta peaks  in the 
distribution function 
 at $\theta_1{=}\Delta_\epsilon$
are essentially broadened, the  
smearing of the gap edges is quite small. Therefore, we shall neglect the corrections due to the high-order harmonics to $P(\theta_1)$ in the rest of the paper.    

\subsection{Effect of the electro-mechanical coupling}

\begin{figure*}[t]
\centerline{\includegraphics[width=0.45\textwidth]{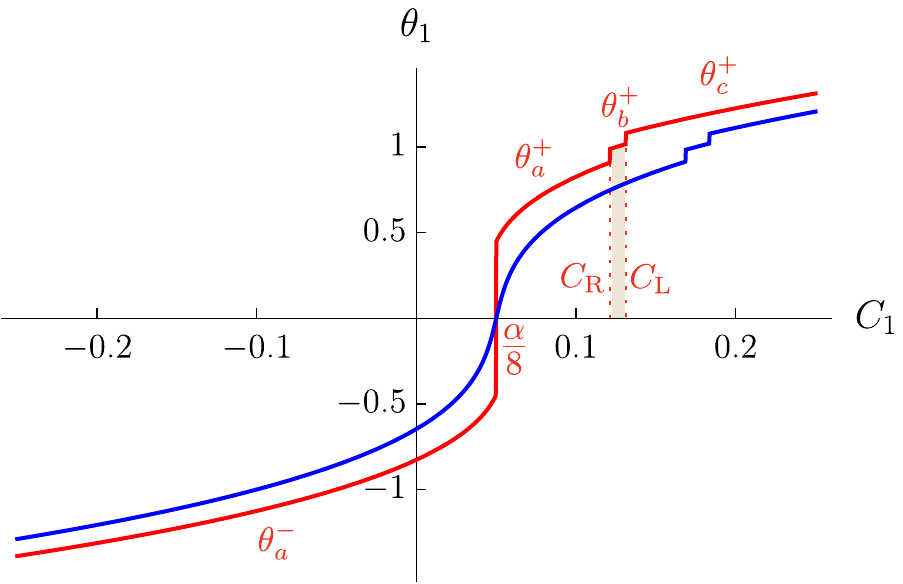}
\hspace{0.05\textwidth}
\includegraphics[width=0.45\textwidth]{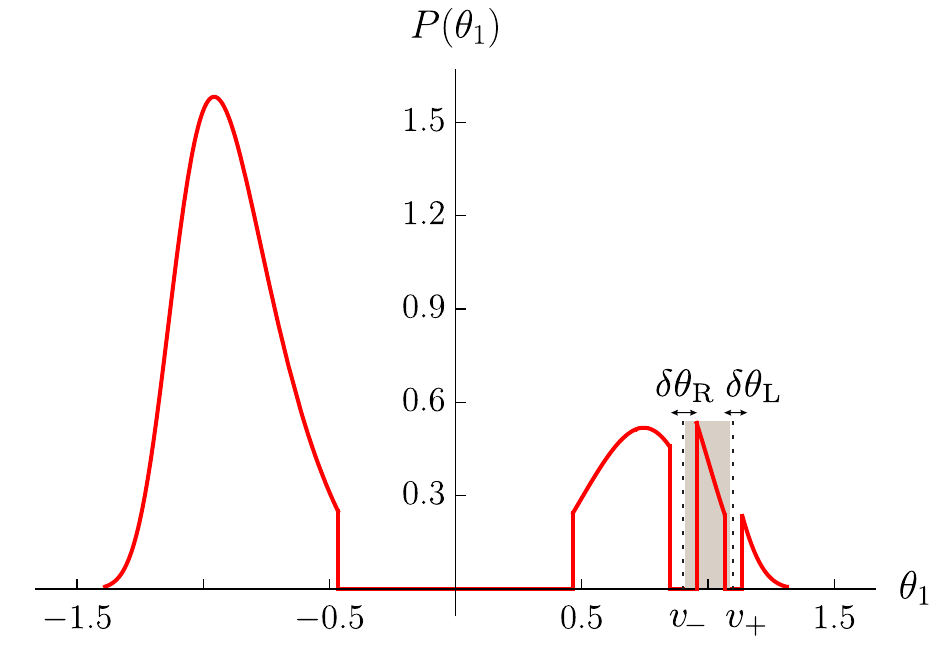}}
\caption{\label{fig:5}Left: Dependence of $\theta_1$ corresponding to the absolute minimum of the energy, on disorder strength  $C_1$ for $\epsilon{=}{\pm} 0.2$ (red and blue) and $\alpha{=}0.4,\;v_g{=}1,\;v{=}0.2.$ 
 Dependence consists of several parts described by functions $\theta_a^-(C_1),~ \theta_a^+(C_1),~\theta_b^+(C_1),$ and $\theta_c^+(C_1).$  Large jump occurs at $C_1{=}\alpha/8,$ while two small mini-jumps occur at the values $C_1{=}C_{\rm R}$ and $C_1{=}C_{\rm L},$  corresponding  to the intersection of the point $(v_g,v).$  by  the  right and left boundaries of  current carrying 
triangle, respectively.    
Right:  Bending angle distribution function
for the curvature  variance  $\Delta_1{=}5\cdot 10^{-3}$ and $\epsilon{=}0.2$.  Grey area in both panels corresponds to non-zero current. Voltages $v_-$ and $v_+$ are situated in the centers of mini-gaps.}
\end{figure*}

As we demonstrated above, equation
$f(\theta_1){=}0$  has several solutions belonging to the manifold    $\theta_s^\pm (C_1) $ (see  Fig.~\ref{f-nd-new}).
 These solutions and corresponding energies 
depend on  $C_1.$   
 Since function $f(\theta_1)$ has jumps at points $\theta_1{=}v_-$ and $\theta_1{=}v_+,$ the function $\theta_1(C_1), $ which, by definition, represents the solution with the minimal energy, might have  mini-jumps in addition to a large jump existing in the case $\alpha{=}0.$  
  Existence  of these mini-jumps    as well as  their  positions 
  and position   of the large jump
  depend on  $v_g, v,$ and on the value of $\alpha.$ 
  For illustration, in  Fig. \ref{fig:5} we  plotted $\theta_1(C_1)$ and corresponding distribution function, 
  \begin{equation}
\label{PthetaC1}
    {P}(\theta_1) =  \bigl[{d\theta_1 (C_1)}/{dC_1}\bigr]^{-1}    \mathcal P(C_1)\Bigl|_{C_1=C_1(\theta_1)} 
\end{equation}
    (here $C_1(\theta_1)$  is the inverse function for  $\theta_1(C_1)$), 
    for the case $v{>}\overline{v}_{\rm Th}(v_g) $ and relatively large $v_g$  
     such that  
     $C_{\rm R}{>}\alpha/8.$
 
As seen from this figure,
there are three major  effects of the electromechanical coupling: (i) the curve $\theta_1(C_1)$ becomes asymmetric with respect to the inversion: $\theta_1\mapsto -\theta_1;$ (ii) several vertical mini-jumps  form on this dependence in addition to the large jump; 
(iii) large jump  between  values of the bending angle  close to $-\sqrt \epsilon$ 
and $\sqrt \epsilon$  shifts from 
$C_1{=}0$  to the point  $C_1{=}\alpha/8$ (this specific value   depends on a position of point in the $(v_g,v)$ plane).

Let us discuss these properties in more detail. First, we notice that  $W^+_s{-}W^-_s$ change sign when $C_s{=}0$ (in particular, for  $|C_s|{\ll} \epsilon^{3/2},$ we get 
$W^+_s{-}W^-_s {\approx}{-}16 C_s \sqrt \epsilon$).   Since, by assumption, $C_{\rm R}{>}\alpha/8,$  large jump occurs with increasing $C_1$ at the point $C_a{=}0,$ i.e. at $C_1{=}\alpha/8,$ where  $W^+_a{=}W^-_a$  and, consequently, there happens a jump  from  $\theta_a^-$ to $\theta_a^+.$ With further increase of $C_1$  it  reach the value 
$C_1{=}C_{\rm R},$ where  solution     jump into current-carrying 
triangle
$\theta_a^+ {\to} \theta_b^+,$ and further jumps out the 
triangle, $\theta_b^+ {\to}  \theta_c^+$  at $C_1{=}C_{\rm L}.$ The values of $C_{\rm L,R}$ can be found from Eqs.~\eqref{v0-small-C1}, \eqref{vg0-small-C1} and \eqref{C-RL}:
\be
C_{\rm R}{\approx } \frac{3 \alpha}{32} {+}\frac{\epsilon}{4} \left( v_- {-} \sqrt{\epsilon} \right),
\quad
C_{\rm L}{\approx} \frac{ \alpha}{32} {+}\frac{\epsilon}{4} \left( v_+ {+} \sqrt{\epsilon} \right).
\ee
The averaged current is given by 
Eq.~\eqref{eq:av:cur},  while the amplitudes of jumps read:
$\delta \theta_{\rm R}=
\theta_b^+(C_{\rm R})-\theta_a^+ (C_{\rm R}),$ 
  $\delta \theta_{\rm L}=
\theta_c^+(C_{\rm L})-\theta_b^+ (C_{\rm L}). $
For small $\alpha ,$ obeying Eq.~\eqref{small-C1-alpha},  we obtain
\be
\delta \theta_{\rm R} \approx \delta \theta_{\rm L} \approx \frac{\alpha}{4\epsilon}. 
\ee

As seen from  Fig.~\ref{fig:5},   probability that the  bending angle  is close to the value 
${-}\sqrt \epsilon$ is larger than the probability    that $\theta_1{\approx}\sqrt \epsilon.$ The probability imbalance increases with decreasing $\Delta_1.$   For  $ \sqrt \Delta_1 {\ll} \alpha,$ the bending angle with exponential  precision is concentrated near  ${-}\sqrt \epsilon.$

Let us now discuss   the  physics behind  
the additional  random-curvature-induced  plateau in the $I-V$ curve. 
The key point is the  competition between two symmetry breaking mechanisms: 
electro-mechanical coupling and  random curvature. 
Remarkably, the limits $\alpha{\to}0$  and $\Delta_1{\to}0$ do not commute.  Indeed, for $\alpha{\equiv}0$  and  variance tending to zero, $\Delta_1{\to}0,$  energy  $W(\theta_1)$
has 
two degenerate minima for 
  $\epsilon{>}0$.
Thus, distribution function is given by the sum of two delta-functions: 
  $$P_{\Delta}={P}(\theta_1)_{\Delta_1{\to}0 ,\alpha{=}0} {\simeq} \frac{1}{2} 
 \left[\delta(\theta_1{+}  
     \sqrt {\epsilon}) + \delta(\theta_1{-}  
     \sqrt {\epsilon})\right].$$
     Although a weak disorder slightly broadens the delta functions, the symmetry ${P}_{\Delta}(\theta_1){=} {P}_{\Delta}({-}\theta_1)$ remains intact.
In the considered limit, $\Delta_1{\to}0$ and $\alpha{=}0$, expression for average current     reads $$\langle I\rangle{=} \frac{1}{2} \int \limits_{v_-}^{v_+} P_{\Delta} (\theta_1) d\theta_1.$$
    One can easily check that this expression       contains   
   an    intermediate plateau  with $\langle I \rangle=1/4$ in a wide interval of $v$  in addition to  the plateau with $\langle I \rangle=1/2$     for  $v \to \infty.$         By contrast,   
   setting $\Delta_1 {=} 0$ first and, then, sending  $\alpha {\to} 0$,
    we find asymmetrical distribution function,
  $$P_{\alpha}={P}_{\Delta_1{=}0,\alpha{\to} 0}(\theta_1) = \delta(\theta_1{+}\sqrt \epsilon ).$$  Corresponding expression for the average current,  
 $$ \langle I\rangle{=} \frac{1}{2} \int \limits_{v_-}^{v_+} P_{\alpha} (\theta_1) d \theta_1,$$
   has only  a single  plateau with $\langle I \rangle=1/2$ for $v\to \infty.$
          The competition between two mechanisms is  
         controlled by the parameter $ \sqrt{\Delta_1}/\alpha$ (see also Appendix \ref{simple-model}).
                      An additional step in the $I-V$ curve appears when this parameter becomes large: $ \sqrt{\Delta_1}/\alpha \gg 1.$    Remarkably, 
                   the current can be  
                    enhanced by disorder contrary to naive expectations (see the bottom panels in  Fig.~\ref{fig:6} and Fig.~\ref{Fig-step-simplified}b).

\section{Discussions and conclusions\label{Sec:Disc}} 
In this paper, we predict curvature-induced enhancement of elastic blockade.  
The maximal 
position of the tip of the current-carrying triangle  is given by
$eV_0^{\rm max}{=}(a_1 E L/\pi) v_0(C_b=0) $ [see Eqs.~\eqref{vvg-dimen} and \eqref{v0-small-C1}].    Hence,
\be e V_{0}^{\rm max} 
\sim e E L\sqrt{\epsilon}. 
\label{Vth-estimate}
\ee 
For $E{\simeq} 10$ kV/cm, $L{\simeq} 0.1\, \mu$m and $\epsilon{\sim}1$ 
it gives $e V_{0}^{\rm max} {\sim} 0.1 $ eV that is larger than the charging energy $E_{\rm c} {\sim} 10$ meV. 
We note that enhanced bias-voltage threshold of the order of $eV_0^{\rm max}$  exists for a sufficiently  weak built-in curvature, in a narrow  window,   $0{<}C_1{<}\alpha/8$. Away from  this window the bias-voltage threshold is parametrically reduced (for $\alpha{\ll}\epsilon^{3/2}$) and      becomes of the order of the threshold in the clean case~\cite{Weick2010, Weick2011,Micchi2015}. From Eq.~\eqref{Vth-C=0}, we find $v_0^{\rm cl}{=}\alpha/4 \epsilon,$ so that   
\begin{equation}
    e V_{0}^{\rm cl} \sim \frac{\alpha}{\epsilon} e E L \ll e V_{0}^{\rm max}. 
\label{eq:Vth:clean}
\end{equation}
Eqs.~\eqref{Vth-estimate} and \eqref{eq:Vth:clean} imply that bias threshold voltage very sharply depends on disorder: it changes by a large factor $\epsilon^{3/2}/\alpha$ in a very narrow region of small $C_1.$     
For the experiments of  Ref.~\cite{Fan2005} we can  roughly estimate absolute value of curvature    $|C_1|\lesssim 0.1{\div}1.$  
For this  estimate we used Eq.~\eqref{theta-lin-C1} with  $F{=}0$ (i.e. $\epsilon{=}{-}8$)  and values of  $\theta_1^{\rm built-in}$ estimated from images of bent nanotubes in Ref.~\cite{Fan2005}. We also assumed that in equilibrium $C_1 {\sim} \theta_1^{\rm built-in } $ (this estimate follows from  Eq.~\eqref{eq-for-theta} for $F{=}0$ and small $\alpha$).   Although the microscopic model of built-in curvature  is absent so far, one may expect that disorder-induced bending angle is not very sensitive to  length of the nanotube. 
Due to smallness  of $\alpha,$ the above estimates imply that  built-in curvature might dominate electron transport through a nanotube-based SET for typical values of experimental parameters. 
It is worth noting that the   threshold voltage given by  Eq.~\eqref{Vth-estimate}  does not depend on the curvature  but   increases with length. Therefore,  for long nanotubes used in Ref.~\cite{Fan2005}, the effect of elastic blockade, which we discuss in the current work, is even higher and value $eV_0^{\rm max}$ is even larger compared to the  estimate presented above. 
A  more detailed comparison with  experiment  requires development of the microscopical  theory of the elastic disorder, and therefore is out of scope of the current work devoted to development of the phenomenological approach.

In the previous sections we assumed that $T{=}0.$
Let us now  discuss the effect of nonzero temperature. For a clean SET  these effects were analyzed in Refs.~\cite{Weick2010,Weick2011,Micchi2015}. The current  depends on $T$  due to at least two reasons. At first, temperature enters the Fermi distribution functions in the expressions \eqref{eq:ndY} and \eqref{J} for the average excess electron number and the current, respectively. 
Secondly, there are  thermal fluctuations of the generalized coordinate $Y$, i.e. $\theta_1$. 
In the fundamental mode approximation these thermal fluctuations of the bending angle are described by the Gibbs factor  
$\exp[{-}\mathcal E_{\rm eff}(\theta_1)/T]{=} \exp({-}W_{\rm eff}(\theta_1)/\tilde{T})$ where $W_{\rm eff}(\theta_1)$ 
and $\mathcal E_{\rm eff}(\theta_1)$  are obtained from  Eqs.~\eqref{W-theta1} and \eqref{E-W} by replacement  [see Eqs. (19) and (25) of Ref.~\cite{Micchi2015}]
$w_d {\to} w_{d,{\rm eff} }(\theta_1){=}({\tilde{T}}/{2})\sum_{\lambda{=}\pm} \ln \left[{f_{\rm F}(\alpha(v_\lambda{-}\theta_1))}/{f_{\rm F}(\alpha v_\lambda)}\right]$. 
 The effective dimensionless  temperature is defined as 
\begin{equation}
\tilde T = \frac{16 L T}{\pi^2 \kappa}.
\end{equation}

As usual, a non-zero temperature 
results in 
smearing of the bias-voltage threshold in the current. 
For a fixed curvature $C_1$, 
our results for the current are valid provided temperature is much smaller than the threshold voltage, 
\begin{equation}
    T\ll e V_{0}(C_1) .
    \label{eq:T:V0}
\end{equation}
Here voltage  $V_{0}(C_1){=}(a_1/\pi) E L v_0(C_1)$, cf. Eq.~\eqref{vth-C1},
 is  a very sharp function of curvature  within the window $0{<}C_1{<}\alpha/8,$  changing 
 within the interval  $V_0^{\rm cl}{<}V_{0}(C_1){\leqslant}V_0^{\rm max}.$     
As follows from the estimates presented above,  $T$ is always smaller than
$V_0^{\rm max}$ up to the room temperatures but can be much larger than  $V_0^{\rm cl}$.
  Hence,   
  the effect of temperature is negligible 
  for $0{<}C_1{<}\alpha/8,$  but it can 
  wash out the bias-voltage threshold for $C_1$ outside this interval. 

\begin{figure}[t]
\includegraphics[width=0.95\columnwidth]{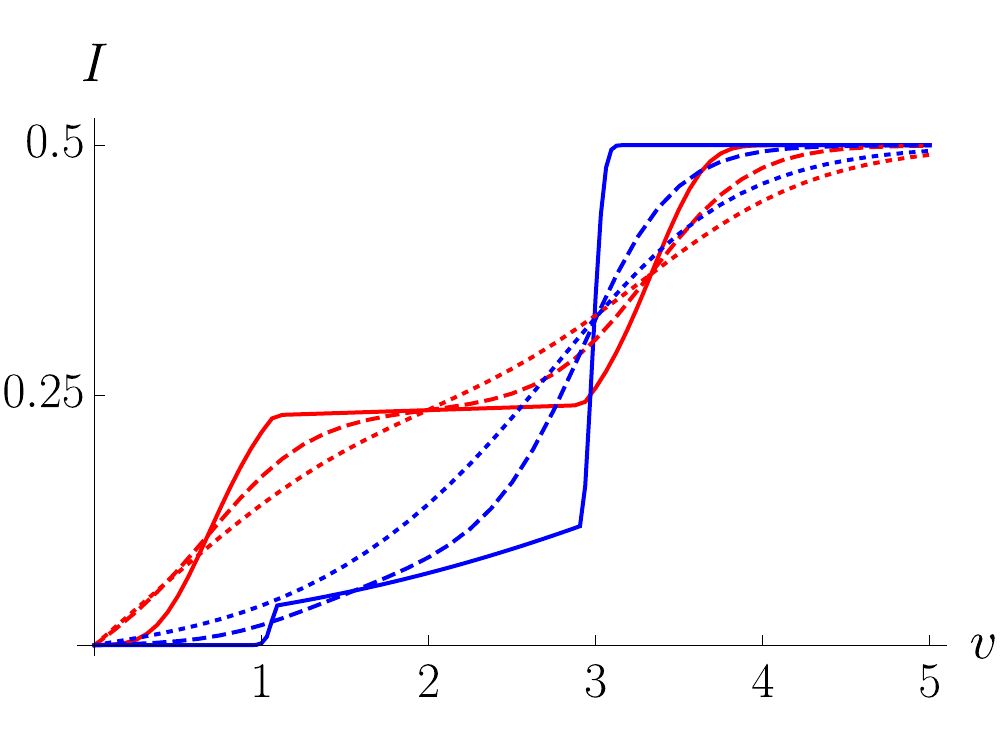}
\caption{
\label{fig:T} 
The current-voltage characteristics for $v_{\rm g}{=}1,$
$\alpha{=}0.02$  and  different values of disorder  variance   $\Delta_1{=}8\cdot 10^{-4}$ and $\Delta_1{=}3\cdot 10^{-6}$ (red and blue correspondingly) for $\tilde T{=} 0,\;2.5 \cdot 10^{-3},\; 5 \cdot 10^{-3}$ (solid, dashed, dotted curves, respectively).}
\end{figure}

In the case of a randomly distributed curvature the enhanced bias-voltage threshold \eqref{Vth-estimate} occurs near zero gate voltage, $V_g {\simeq} 0$. Away from the elastic blockade peak the bias-voltage threshold is suppressed down to $V_{0}^{\rm cl}$, cf. Eq.\eqref{eq:Vth:clean}. 
The  thermal smearing 
is not important at temperatures $T{\ll} e \bar{V}_{\rm Th}(V_g)$ where $\bar{V}_{\rm Th}(V_g){=} (a_1/\pi) E L  \bar{v}_{\rm Th}(v_g)$, cf. Eq. \eqref{vth-vg}. 
\color{black}
For a random curvature distributed continuously, the average current is given as the average over $P(C_1)$ and the Gibbs weight $\exp(-W_{\rm eff}(\theta_1)/\tilde{T})$.  The result of the corresponding numerical  computation is shown on the Fig. \ref{fig:T}. It illustrates 
blurring of the disorder-induced step in the $I-V$ curve at intermediate voltages by a finite temperature.

One of the most interesting problems to be solved in future is to study effect of disorder on high-quality resonances observed in Refs.~\cite{Bonis2018,Barnard2019,Erbil2020,Wang2021}.  

To summarize, we considered transport properties of a 
SET based on elastic  nanotube with a 
 regular or random  built-in
 curvature 
in the  strong Coulomb blockade 
 regime.  We 
 demonstrated  that close to the buckling transition, the 
   $I$-$V$ curve  of the transistor
is extremely sensitive to  such
curvature.
 
 Most importantly,  we predict the following
 \begin{itemize}
 \item[$\bullet$] for fixed  built-in curvature:  a giant curvature-induced enhancement of the bias-voltage threshold \eqref{Vth-estimate}
below which the current is exactly zero; 

 \item[$\bullet$]  for random curvature: the existence of an additional intermediate  curvature-induced plateau  in average  $I-V$ curve, see Fig. \ref{fig:6}.  
 \end{itemize}
Our predictions can be tested experimentally in systems similar to those studied in Refs. \cite{Bonis2018,Barnard2019,Erbil2020,Wang2021}. 
 \begin{acknowledgements}
 We thank I. Gornyi  
 and A. Shevyrin 
 for useful comments and discussions. The
work was funded in part by  Russian Foundation for Basic Research (grant No. 20-52-12019) -- Deutsche Forschungsgemeinschaft (grant No. SCHM 1031/12-1) cooperation.  VK was partially supported by  
 IRAP  
Programme  of  the Foundation   for   Polish
   Science   (grant   MAB/2018/9, project CENTERA). 
\end{acknowledgements}

\appendix
\section{Effect of higher harmonics of the curvature}
\label{high-harm}

In this Appendix, we estimate effect of higher modes and demonstrate that this effect is small. We will focus on modification of the  distribution function of  
  $\theta_1$ caused by higher harmonics, neglecting electro-mechanical coupling. Generalization for the case $\alpha{\neq}0$ is  straightforward but  quite cumbersome and  does not change our main conclusion about irrelevance of the higher order terms.      

In order to find the equation for $\theta_1,$ we perform an expansion of $\sin \theta$ to the third order,  $\sin \theta{\approx} \theta{-}\theta^3/6$. Next, we  
use  expansion~\eqref{expansion} 
and, then,
project the result onto the first harmonic, $n{=}1$. We keep terms up to the  $\theta_n^3$.
After some algebra, we arrive at the following cubic equation,
 \be
     \theta_1^3 -  (\epsilon -\delta \epsilon) \theta_1  + \theta_1^2 \theta_3  - 8 C_1=0 ,
 \label{n=1-new}
 \ee 
  where 
    \be
         \delta \epsilon= 2 \sum_{n=2}^\infty  \theta_n \bigl (  \theta_n+ \theta_{n+2}      \bigr ).
 \label{e-shift-new}
 \ee
Here  harmonics  with $n{>}1$ for $|\epsilon| {\ll} 1$ can be calculated within linear approximation, so we obtain 
\be
\theta_n \approx C_n \frac{q_n^2 }{q_n^2-q_1^2},
\quad \text{for} \quad n=2,3,\dots
\label{n>2}
\ee
They turn out to be finite at the instability threshold, $\epsilon{=}0.$  Consequently, the first harmonics dominate over higher-order ones. Hence, we can consider effect of higher harmonics perturbatively. 
As it follows from  Eq.~\eqref{n>2}, the amplitudes $\theta_{n}$ with $n{\geqslant} 2$ in  Eqs.~\eqref{n=1-new} and \eqref{e-shift-new}  are fixed by the disorder.
Therefore, the problem is reduced to the  analysis of the closed equation~\eqref{n=1-new}  for the  most singular zero mode with the amplitude $\theta_1$.  Just as in the case of single-mode approximation,  for a fixed realization of disorder there are one or three real solutions of Eq.~\eqref{n=1-new} for $\theta_1$.

Next, we discuss  the effect of higher harmonics on the distribution function. We shall perform calculations in two steps. First, we average over fundamental harmonic $C_1$ for fixed  $C_{n>1}.$ Next, we do averaging  over higher harmonics.  Let us 
 introduce a new variable,
\be  \theta_{1,{\rm eff}}= \theta_1 +\frac{\theta_3}{3}=\theta_1+ \frac{3 C_3}{8} . \label{theta-new}
\ee
Here we used Eq.~\eqref{n>2}.  Substituting Eq.~\eqref{theta-new} into Eq.~\eqref{n=1-new}, we get
\be
\theta_{1,{\rm eff}}^3- \epsilon_{\rm eff}\theta_{1,{\rm eff}} = 8\tilde C_1, \label{theta1-new}
\ee
where
\be
\begin{aligned}
&\epsilon_{\rm eff}=
\epsilon -\delta \epsilon +\frac{27}{64}C_3^2, 
\\ 
&\tilde C_1=
C_1-\frac{1}{4}\left(\frac{3C_3}{8}  \right)^3-\frac{3C_3}{64}(\epsilon-\delta \epsilon).
\end{aligned}
\ee
Equation \eqref{theta1-new} is equivalent to  Eq.~\eqref{theta1-C1} up to a change of variables.
Hence, by  using calculations presented in the main text, we derive equation analogous to Eq.~\eqref{P-C3=0}
\begin{gather}
P(\theta_1)=
\exp\left[-\frac{(\theta_1^3-\epsilon \theta_1)^2}{128 \Delta_1}\right]
\left \langle\frac{3 \theta_1^2-\epsilon +\delta \epsilon +9 \theta_1 C_3/4 }{8\sqrt{2\pi \Delta_1}} \right.
\notag \\
\times 
\left. 
\left[\Theta\left( \theta_1+\frac{3 C_3}{8}-\Delta_{\epsilon_{\rm eff}}\right)\!+\!\Theta\left(- \theta_1 -\frac{3 C_3}{8}-\Delta_{\epsilon_{\rm eff}}\right)\right]\right\rangle
\label{P-C3-neq0} 
\end{gather}
Here we calculated Jacobian  by using Eq.~\eqref{n=1-new}. We also  took into account that  corrections due to high harmonics are small and  can be fully neglected in the argument of the  the exponent.   The averaging in Eq.~\eqref{P-C3-neq0}   is taken over  $C_{n>1}.$  

Further consideration depends on relation between  $ \Delta_1 $ and $\epsilon.$  Below, we consider  limits of both weak and strong disorder.

\subsection{Weak disorder, $\Delta_1 {\ll} |\epsilon| {\ll} 1$}

In this case, one can neglect disorder everywhere except the arguments of the step function, where we can also neglect the difference between  $\epsilon_{\rm eff}$ and $\epsilon.$ Then disorder averaging is reduced  to averaging of the step functions:
\be
\left \langle\Theta\left(\!\!\pm\left[ \theta_1+\frac{3 C_3}{8} \right]-\Delta_\epsilon \right)\right \rangle_{C_{3}}\!\!\!\!\!=\frac{1}{2} {\rm erfc}\left[\frac{8(\Delta_\epsilon \mp \theta_1)}{\sqrt{2\Delta_1}}\right].
\ee
Here we 
used $\Delta_3{=}\Delta_1/9$ (cf. Eq.~\eqref{corr-Cn}).
The distribution function becomes 
\begin{align}
\label{weak0}
    &P(\theta_1) = \frac { 3 \theta_1^2 - \epsilon  } { 16 \sqrt { 2 \pi \Delta_1 } } \exp\left[-\frac  { \left( \theta_1^3 - \epsilon \theta_1 \right)^2 } { 128 \Delta_1 } \right] 
    \\
    \nonumber
    &\times 
        \left\{{\rm erfc} \left[\frac { 8 \left( \Delta_\epsilon  -  \theta_1 \right) } { \sqrt { 2 \Delta_1 } }\right] +{\rm erfc} \left[\frac { 8 \left(  \Delta_\epsilon  +  \theta_1 \right) } { \sqrt { 2 \Delta_1 } }\right]\right\}. 
\end{align}
Hence, for weak disorder, the only effect of the higher harmonics is blurring  the step-like edge of the  gap (for $\epsilon{>}0$)  over a very narrow  window $\delta\theta_1 {\sim} \sqrt \Delta_1{\ll} \sqrt \epsilon.$

\subsection{Strong disorder, $\epsilon {\ll} \Delta_1 {\ll}  1$}

For strong disorder, one can neglect  contribution of $\epsilon$ to $\epsilon_{\rm eff}.$ Then,
$\epsilon_{\rm eff} {\approx} 
 {-}\delta \epsilon {+} 27C_3^2/64$.
Then using the relation  
\begin{gather}
2\sum\limits_{k=2}^\infty \left (\theta_k^2+\theta_k \theta_{k+2}\right ) =
\Bigl[ \theta_2^2+ (\theta_2+\theta_4)^2 + (\theta_4+\theta_6)^2+\dots \Bigr ] \notag \\
 + \Bigl [\theta_3^2+(\theta_3+\theta_5)^2+(\theta_5+\theta_7)^2 + \dots \Bigr ] ,
      	\end{gather}
we prove that $\epsilon_{\rm eff} {<}0.$ Hence, $\Delta_{\epsilon_{\rm eff}}{=}0,$ so that  the sum of two step functions in Eq.~\eqref{P-C3-neq0} equals to $1.$ Hence, we only need to average $\delta \epsilon$ in Eq.~\eqref{P-C3-neq0}. Straightforward calculation yields
\be
\big\langle\delta \epsilon \big\rangle_{C_{n>1}}=2\Delta_1 \sum \limits_{n=2}^{n=\infty}\frac{n^2}{(n^2-1)^2}=\Delta_1\frac{3+4\pi^2}{24} .
\ee
Here, we used Eqs.~\eqref{corr-Cn} and \eqref{e-shift-new}. 
Substituting $\big\langle\delta \epsilon \big\rangle_{C_{n>1}}$  into Eq.~\eqref{P-C3-neq0},  we obtain distribution function for  $|\epsilon|{\ll}\Delta_1{\ll}1$:
\begin{gather}
P(\theta_1)\! =\! \frac{3 \theta_1^2
\!+\! {(3+4\pi^2)\Delta_1}/{24} }{8\sqrt{2\pi \Delta_1}}
    \exp\left [-\frac{\theta_1^2\left( \theta_1^2 - \epsilon\right)^2}{128\Delta_1}\!\right ].
\label{P-strong}
\end{gather}

\subsection{Interpolation  formula and gap formation}

        One  can easily find  a 
        cross-over function that interpolates between  weak and strong disorder  limits. We note that 
                both $\Delta_1$ and $\epsilon$ are still assumed to be smaller than unity. The cross-over function reads
        \begin{align}
   P(\theta_1)  & \simeq \frac {  3 \theta_1^2 - \epsilon + (3+4\pi^2)\Delta_1/24   } { 16 \sqrt { 2 \pi \Delta_1 } }  \exp \left[ - \frac { \left( \theta_1^3 - \epsilon \theta_1 \right)^2 } { 128 \Delta_1 } \right] 
   \notag \\
   & \times \left[ {\rm erfc} \left (\frac { 8 \left( \Delta_{ \epsilon } - \theta_1 \right) } { \sqrt { 2 \Delta_1 } }\right ) + {\rm erfc}\left(  \frac { 8 \left( \Delta_{ \epsilon } + \theta_1 \right) } { \sqrt { 2 \Delta_1 } }\right ) \right].
\label{cross-P}
\end{align}

\begin{figure}[t]
\includegraphics[width=.95\columnwidth]
{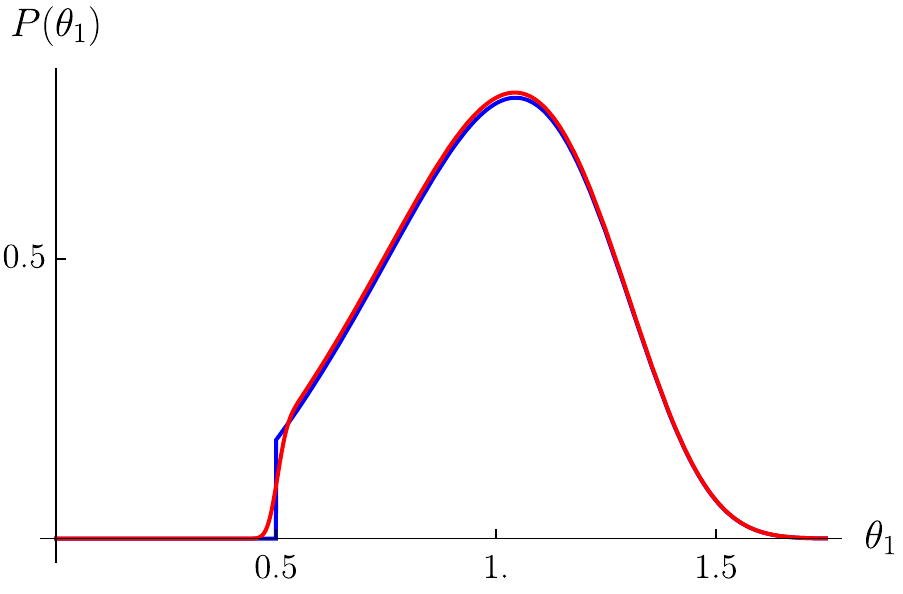}
\caption{\label{cross-P-f}
Distribution function calculated with the use of Eqs.~\eqref{P-C3=0} (blue) and  \eqref{cross-P} (red)  for 
$\Delta_1{=}0.02$ and $\epsilon{=}0.25$.}
\end{figure}

Equation~\eqref{cross-P} describes formation of the gap in the distribution function.   Due to coupling   with higher harmonics, the edge of the gap is not infinitely sharp.  Formally this is due to presence of $\rm ercf-$functions in  Eq.~\eqref{cross-P} instead of $\Theta[\theta_1^2-\epsilon]$ in Eq.~\eqref{P-C3=0}. Deep inside the gap, say for $\theta_1=0,$  perturbative expansion over harmonics fails and  Eq.~\eqref{cross-P} becomes invalid. Correct  calculation of  exponentially small tails of the distribution function in the middle of the gap can be done by using optimal fluctuation method. Such calculation is out of scope of the current work. We also notice that comparison of Eqs. ~\eqref{cross-P}  and \eqref{P-C3=0} shown  in Fig.~\ref{cross-P-f} shows that difference between distribution functions calculated with and without account of high harmonics is quite small.

\begin{figure}[t]
\includegraphics[width=\columnwidth]{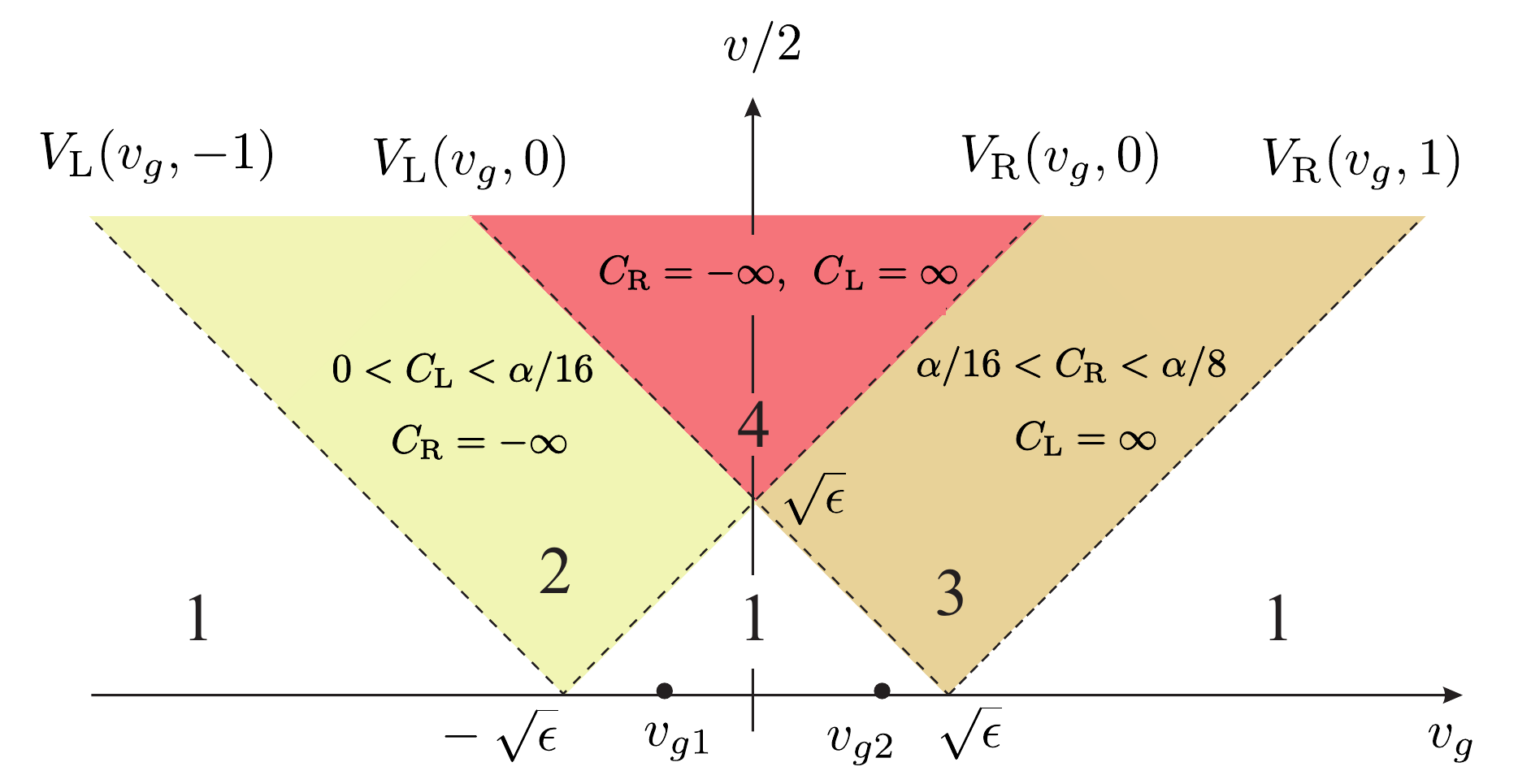}
\caption{ \label{Fig-simplified} 
Phase diagram of the simplified model. The averaged current is equal to zero in the region $1$. It is equal  to $1/2$ in the region $4.$ Dashed lines show positions of the  right and left boundaries of the current-carrying 
triangle for $\xi{=}{-}1,0,1.$ 
Voltage-current curves for $v_{g}{=}v_{g1}$ and $v_g{=}v_{g2}$ are plotted in Fig.~\eqref{Fig-step-simplified} }
\end{figure}

\section{ Simplified model}\label{simple-model}

Here, we  discuss in more detail  averaged current obtained within simplified model, when both $C_1$ and $\alpha $  tend to zero in such a way  that the ratio  $C_1/\alpha$ remains finite, cf. Eq.~\eqref{xi-approximation} and $\epsilon{>}0.$  Approximation    \eqref{xi-approximation} allows us to find exactly values of $C_{\rm L}$ and $C_{\rm  R}$ entering Eq.~\eqref{eq:av:cur}, and, consequently, to find simple analytical equation for the curvature-averaged $I$-$V$ curve.

\begin{figure}[t]
\includegraphics[width=0.9\columnwidth]
{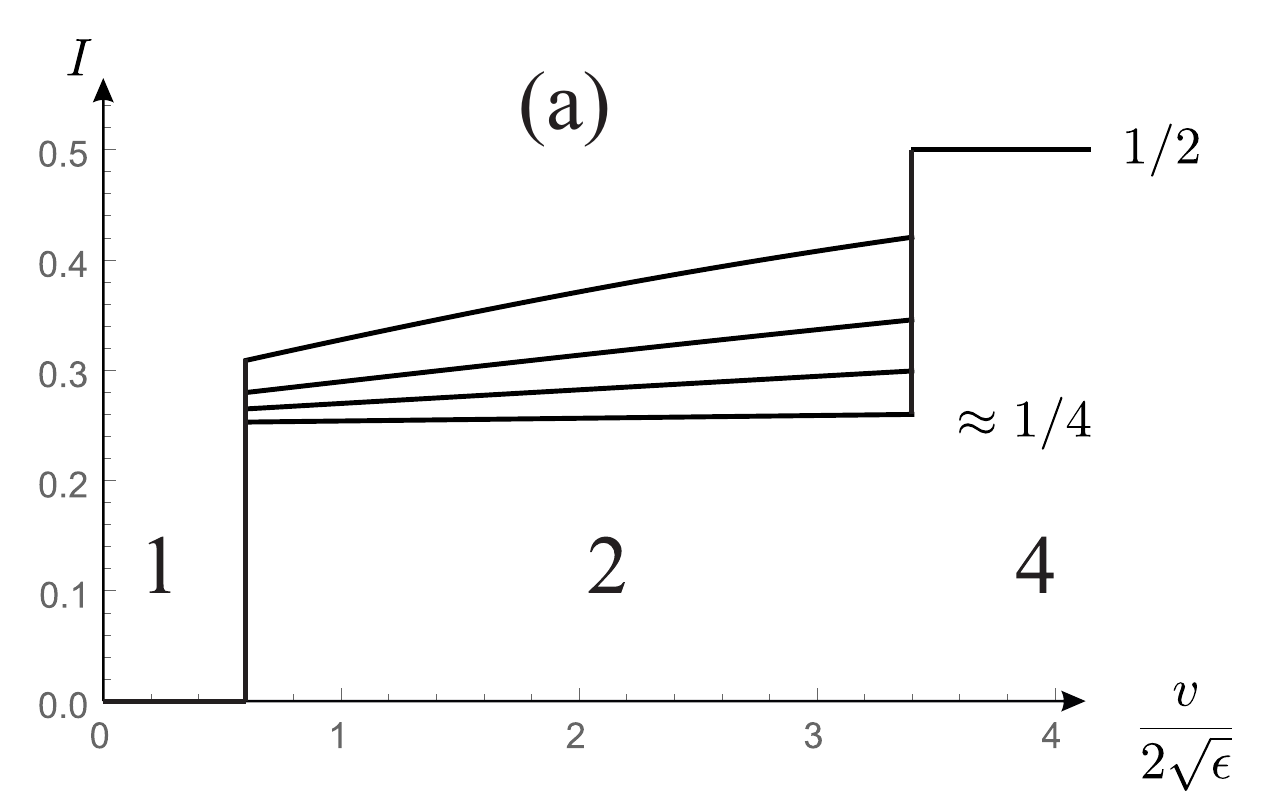}
\includegraphics[width=0.9\columnwidth]
{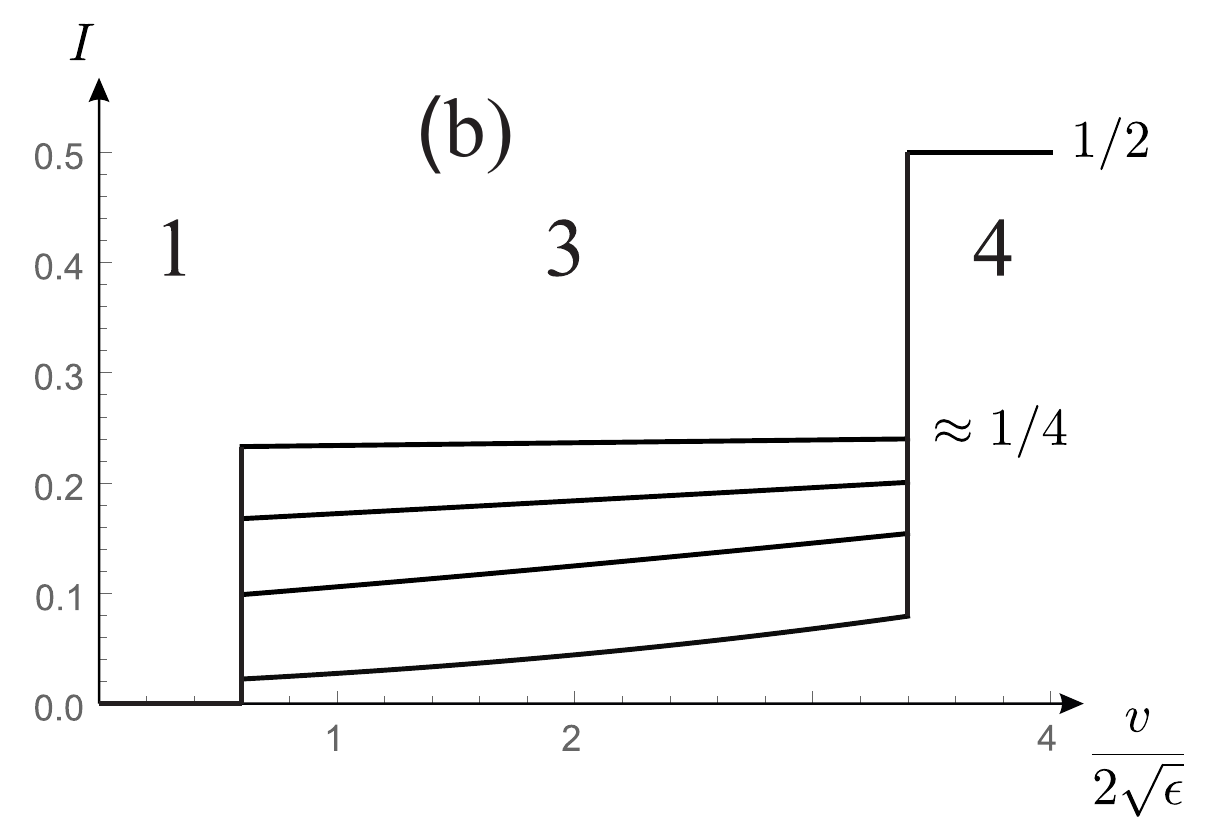}
\caption{\label{Fig-step-simplified} 
Voltage-current curves calculated within simplified model for two values $v_g,$ corresponding to lines $v_{g}{=}v_{g1}{=}{-}0.7 \sqrt \epsilon$ (a)   and $v_g{=}v_{g2}{=}0.7 \sqrt \epsilon$ (b) (see also Fig.~\ref{Fig-simplified}) for $\alpha{=}0.016$ and   different   values of variance $\Delta_1$:  $\sqrt \Delta_1{=}0.001,~0.002,~0.004, ~0.2,$ which increases from  bottom to top in panel (a) and from  top to bottom in panel (b). In both cases, for large $\Delta_1,$ an intermediate step with $I=1/4$ is formed. }
\end{figure}

Using Eq.~\eqref{xi-approximation}, one can easily find how left and right boundaries of the current-carrying 
triangle move with changing $\xi$ in the interval $-1{<}\xi{<}1$: 
\begin{align}
&v_{\rm L} (v_g,\xi){=}2 (\sqrt \epsilon -v_g) {+} 4 \sqrt \epsilon ~ \xi \Theta \left(-\xi\right),
\label{L-xi}
\\
&v_{\rm R} (v_g,\xi){=}2 (\sqrt \epsilon +v_g) {-} 4 \sqrt \epsilon ~\xi \Theta \left(\xi\right).
\label{R-xi}
\end{align}

For $ \xi{<}{-}1,$ the 
triangle does not move and is limited by lines $v_{\rm L}(v_g,{-}1)$ and  $v_{\rm R}(v_g,0).$  
For $\xi{>}1,$ the triangle
is also stationary and is 
bounded by lines
$v_{\rm L} (v_g,0)$ and $v_{\rm R}(v_g,1).$ 

We notice that in the interval $ {-}1{<}\xi{<}0,$  the left boundary 
changes according to Eq.~\eqref{L-xi}, while right boundary remains stationary and is still given by $v_{\rm R}(v_g,0).$  
For $0{<}\xi{<}1,$  left boundary stops and is given by   $v_{\rm L}(v_g,0),$ while right boundary starts to move according to Eq.~\eqref{R-xi}.  This is  very similar to Fig.~\ref{fig:1} with two minor distinctions: in the simplified model the  boundaries  of the current-carrying triangle are parallel  to blue lines and we neglect slow motion of triangle for $C_1<0$ and $C_1> \alpha/8.$

Using Eqs.~\eqref{C-RL}, \eqref{L-xi} and \eqref{R-xi} one can derive $C_{\rm L,R}$ in the regions $1,2,3,4$ of the phase diagram shown in  Fig.~\ref{Fig-simplified}. For any value of  the curvature, points in the region  $1$ are not covered by the current-carrying 
triangle, so that  $C_{\rm L}= C_{\rm R}=\infty$ and $I \equiv 0.$  By contrast,  all points of   the region $4$ belong to the current-carrying cone for any $C_1,$ hence, $I\equiv 1/2.$  

For region $2$ we find  $C_{\rm R}=-\infty,$ and
\be
C_{\rm L}=\frac{\alpha}{64} \frac{v+ 2(\sqrt \epsilon +v_g)}{ \sqrt \epsilon },
\qquad 
0<C_{\rm L}< \frac{\alpha}{16}. 
\ee
Here $C_{\rm L}$ is found from the condition $v{=}v_{\rm L} (v_g, 16 C_{L}/\alpha{-}1 ).$
The averaged current in the region $2$ reads
\be
I^{(2)}=\frac{1}{2} \int \limits_{-\infty}^{C_{\rm L}} P(C_1)dC_1
=\frac{1}{4}\left [1+{\rm erf }\left( \frac{C_{\rm L}}{\sqrt{2 \Delta_1}} \right) \right].  
\label{I2}
\ee

In the region  $3$ we get   $C_{\rm L}=\infty$ and
\be
C_{\rm R}=\frac{\alpha}{64} \frac{ 2(3\sqrt \epsilon +v_g) -v}{ \sqrt \epsilon },
\qquad \frac{\alpha}{16} <C_{\rm R}<\frac{\alpha}{8} . 
\ee
This result for $C_{\rm R}$ is derived from the condition $v{=}v_{\rm R} (v_g, 16 C_{R}/\alpha{-}1 ).$
Averaged current in the region $3$ reads
\be
I^{(3)}=\frac{1}{2} \int \limits^{\infty}_{C_{\rm R}} P(C_1)dC_1
=\frac{1}{4}{\rm erfc }\left( \frac{C_{\rm R}}{\sqrt{2 \Delta_1}} \right) .  
\label{I3}
\ee

Voltage-current curves calculated within simplified model for  two values of the gate voltage  $v_{g}{=}v_{g1}$   and $v_g{=}v_{g2}$ are shown in Fig.~\ref{Fig-step-simplified}. The parameter $\alpha$ is kept fixed whereas  the disorder strength varies.
In both cases, for small driving voltage, in region $1,$ the current is equal to zero, while at very large $v,$ in region $4,$ the current is equal to $1/2.$   In the region $2$ (see Fig.~\ref{Fig-step-simplified}a) the current is given by Eq.~\eqref{I2}. In the region $3$ (see  Fig.~\ref{Fig-step-simplified}b) the current is given by Eq. \eqref{I3}. The  main difference of the simplified model as compared to the exact one is  the presence of sharp jumps  between different regions of the $I-V$ curve. These jump are smoothed in a more accurate model (compare bottom panel of Fig. \ref{fig:6} and Fig.~\ref{Fig-step-simplified}).       
 
\color{black}


\end{document}